\documentclass[12pt, preprint]{aastex}
\usepackage{times}
\newcommand{\etal}{{et al}\/.}
\newcommand{\hh}{^{\rm h}}
\newcommand{\mm}{^{\rm m}}
\begin{document}
\slugcomment{Draft of \today}
\shorttitle{3C\,285 and 3C\,442A}
\shortauthors{M.J.\ Hardcastle \etal}
\title{The interaction between radio lobes and hot gas in the nearby
  radio galaxies 3C\,285 and 3C\,442A}
\author{M.J.\ Hardcastle\altaffilmark{1}, R.P. Kraft\altaffilmark{2},
  D.M. Worrall\altaffilmark{3}, J.H. Croston\altaffilmark{1}, D.A.
  Evans\altaffilmark{2}, M. Birkinshaw\altaffilmark{3}, and S.S. Murray\altaffilmark{2}}
\altaffiltext{1}{School of Physics, Astronomy \& Mathematics, University of
  Hertfordshire, College Lane, Hatfield AL10 9AB, UK}
\altaffiltext{2}{Harvard-Smithsonian Center for Astrophysics, 60 Garden Street, Cambridge, MA~02138, USA}
\altaffiltext{3}{Department of Physics, University of Bristol, Tyndall Avenue,
Bristol BS8 1TL, UK}
\begin{abstract}
We present {\it Chandra} observations of two nearby radio galaxies in
group environments, 3C\,285 and 3C\,442A. The host galaxies of both
sources are involved in mergers with nearby massive galaxies, and the
hot gas in the systems is extended along lines joining the interacting
galaxies. Both sources show strong evidence for interactions between
the radio lobes and the asymmetrical hot gas. We argue that the
structure in the hot gas is independent of the existence of the radio
lobes in these systems, and argue that hot gas shaped by an ongoing
massive galaxy merger may play an important role in the dynamics of
radio lobes in other objects. For 3C\,442A, our observations show that
gas is being driven out of both members of the host interacting galaxy
pair, and the implied constraints on galaxy velocities are consistent
with mildly supersonic motions with respect to the group-scale hot
gas. The previously known filamentary radio structure in the center of
3C\,442A may be a result of the interaction between hot gas expelled
from these galaxies and pre-existing radio-emitting plasma. In
3C\,285, where there is no ongoing galaxy merger, the powerful radio source is
probably having a significant effect on the energetics of the host
group.
\end{abstract}

\section{Introduction}

\subsection{Radio lobes, hot gas, and mergers}

{\it Chandra} and {\it XMM-Newton} observations now routinely allow us
to measure both the inverse-Compton emission from the lobes of radio
galaxies and the thermal brems\-strahlung from the intra-group or
intra-cluster media in which they are embedded
\citep[e.g.,][]{hbch02,cbhw04}. In almost all cases studied to date,
the internal pressures in the radio lobes are comparable to those in
the hot, X-ray emitting phase of the external medium, rather than
greatly exceeding it as had sometimes been thought to be the case
\citep[cf.][]{hw00}. We know of only a few objects --- the best
example being the inner lobes of Centaurus A \citep{kvfj03} --- in
which the internal lobe pressures must be much larger than the
inferred external pressure, although the limited numbers of such
sources found to date is almost certainly affected by selection effects.

In the objects where the pressures of the radio-emitting plasma and
the X-ray-emitting external medium are comparable, the dynamics of the
large-scale radio structures, the lobes and plumes, will be
determined, at least in part, by the properties of the hot gas. Clear
evidence that this is so pre-dates the launch of {\it
Chandra} and {\it XMM-Newton} \citep*[e.g.,][]{hwb98} but much more
detailed studies have been possible in recent years, particularly in
nearby, low-power radio galaxies \citep*[e.g.][]{chbw03,hsw05}. At the
same time, the lobes displace and compress the material that
they expand into, as is evident from observations of `cavities' in the
X-ray-emitting material \citep[e.g.,][]{brmw04}. This means that, when
we observe a radio galaxy and its environment, and see a relationship
between radio-emitting structures and those in the X-ray, it is often
not clear whether what we are seeing is a pre-existing distribution of
hot gas that has shaped the radio source that lies in it, or whether,
on the other hand, we are seeing the {\it effects} of the radio source
on what would otherwise have been a relatively structure-free external
medium.

In a relaxed system that does not host a powerful radio source we do
not expect to find long-lasting complex structure in the hot
X-ray-emitting plasma of a group or cluster atmosphere, since any such
structure will be washed out on a sound-crossing time. However,
complex structures can be created by dynamical processes related to
the group or cluster galaxies. Cluster or group mergers are known to
produce large-scale asymmetries in the ICM/IGM which may affect the
dynamics of radio sources hosted by one of the group/cluster galaxies
\citep[e.g.,][]{kjnh06}. On smaller scales, galaxy-galaxy interactions
and mergers would be expected to give rise to structures in the group
or cluster atmosphere, both because they produce a local gravitational
potential that is not radially symmetrical and because frictional and
tidal processes will give rise to the stripping and heating of gas
from the galaxy \citep*[e.g.,][]{opf04}: the latter process is of
course important if one or both of the merging systems are massive
elliptical galaxies with large pre-existing galaxy-scale hot gas
haloes. This makes it particularly interesting to observe radio
galaxies in which a major merger is taking place. In this paper we
present new {\it Chandra} observations of two nearby radio galaxies,
3C\,285 and 3C\,442A, which meet this selection criterion. We show
that structures in the X-ray-emitting gas, likely related to the
mergers, are almost certainly affecting the dynamics of the lobes of
the radio galaxies, and discuss the implications of this result for
the radio galaxy population as a whole. More detailed discussion of
the dynamics of 3C\,442A in relation to the ongoing galaxy merger can
be found in \citet{wkbh07}.

\subsection{3C\,285}

3C\,285 is a nearby ($z=0.0794$, \citealt{s67}) low-power FRII radio
galaxy, with a 178-MHz luminosity\footnote{Here, and throughout the
paper, we use a concordance cosmology with $H_0 = 70$ km s$^{-1}$
Mpc$^{-1}$, $\Omega_{\rm m} = 0.3$ and $\Omega_\Lambda = 0.7$.} of
$1.5 \times 10^{25}$ W Hz$^{-1}$ sr$^{-1}$, just above the canonical
Fanaroff-Riley \citeyearpar{fr74} break at $\sim 5 \times 10^{24}$ W
Hz$^{-1}$ sr$^{-1}$. At the distance of 3C\,285 1$''$ corresponds to a
projected distance of 1.50 kpc: the luminosity distance is 360 Mpc.
Although 3C\,285 shows jets and hotspots (best seen in the images of
\citealt{vd93}) the jets are relatively bright, and the hotspots
relatively weak, compared to more luminous FRII sources, suggesting
that it is in some sense an intermediate object. 3C\,285's host galaxy
is the brightest member of a small optical group \citep{s72} and is
strongly elongated towards a fainter spiral companion 40$''$ (60 kpc)
to the NNW, suggesting a tidal interaction: an extended filament of
optical emission-line gas connects the two galaxies \citep{bhbv88}.
Several other galaxies with magnitudes that make them likely to be
group members lie within 200 kpc. In {\it HST} imaging
\citep{askm02} 3C\,285's host galaxy can be seen to be strongly
disturbed, with several distinct linear dust features, suggesting a
recent merger with a large, gas-rich galaxy \citep[cf.][]{re00}.
3C\,285 is best known for the association of its radio jet with a
blue, star-forming region, making it one of the few objects to show
direct evidence for jet-induced star formation \citep{vd93}.
\citet*{stc78} claim a detection of an optical counterpart to the E
hotspot, which we discuss further below (\S\ref{opt-hotspot}). 3C\,285
had not been detected in the X-ray until the observations discussed in
the present paper: the {\it Einstein} HRI upper limit quoted by
\citet{fmtl84} shows that the system is not very X-ray luminous (their
limit corresponds to $L_{\rm X} < 2.3 \times 10^{42}$ erg s$^{-1}$
between 0.5 and 3.0 keV within $2'$ in the cosmology we use).

\subsection{3C\,442A}

3C\,442A\footnote{3C\,442A is sometimes referred to as 3C\,442 in the
literature. Early maps of the field \citep[e.g.][]{m69} showed the
presence of two radio sources. One of these sources was extended and
associated with the NGC 7236/7237 galaxy pair; the other, 28 arcmin to
the SSE, was compact, with a flux density around 1 Jy at 1.4 GHz, had
a steep spectrum, and was not associated with any bright galaxy. Its
178-MHz flux density contributes to the measured 3CR flux \citep{v77}.
The designation 3C\,442A was given to the first source (apparently
first used in print by \citealt*{jpr77}; subsequently adopted by
\citealt*{lrl83}) to make it clear that the two were unrelated. The
second source could have become known as 3C\,442B, but is in fact
generally referred to as 4C\,13.83 or PKS 2212+131, and is coincident
with an SDSS quasar with $z=1.90$. Because of its distance from
3C\,442A it is not detected in any of the X-ray datasets discussed in
this paper.} is associated with the interacting galaxy pair NGC 7236/7
at $z=0.027$ \citep{g62}. Its 178-MHz luminosity of $2 \times 10^{24}$
W Hz$^{-1}$ sr$^{-1}$ puts it below the FRI/FRII break. At the
distance of 3C\,442A 1$''$ is 0.54 kpc: the luminosity distance is 118
Mpc. In the radio \citep*{blp81,co91}, 3C\,442A shows a double-lobed
structure with no sign of any jet, although a compact flat-spectrum
radio core coincident with NGC 7237 suggests current AGN activity in
that object. \citeauthor{co91} argue that the compact features in and
close to the lobes are unrelated to the source, which, if true, would
make 3C\,442A a member of the class of `fat doubles' or `relaxed
doubles' \citep[e.g.,][]{l93}, a number of which are found in the 3CRR
sample close to the FRI/FRII luminosity break. 3C\,442A is remarkable
for the detection by \citeauthor{co91} of two filaments of
steep-spectrum radio emission that cross the region containing the
host galaxy, NGC 7237. Optically this galaxy, an elliptical (assumed
to be the host on the basis that it hosts a compact radio source)
appears to be interacting with an S0 neighbour of similar luminosity
to the NE, NGC 7236: the two are separated by 35$''$ (19 kpc) and are
embedded in a common, distorted stellar envelope which also includes
another, smaller elliptical 38$''$ (21 kpc) to the SW of NGC 7237,
denoted NGC 7237C hereafter \citep{bh88}. The measured radial
velocities for NGC 7236 and NGC 7237 are very similar, suggesting an
interaction close to the plane of the sky, which allowed \citet{b88}
to make detailed models of the system. {\it HST} imaging
\citep{mbsw99} shows some weak evidence for dust emission in NGC 7237,
but generally the inner parts of the system appear like an undisturbed
elliptical, in contrast to 3C\,285's host. NGC 7236/7 lies at the
center of a group of galaxies \citep{co91}. In the X-ray, \citet{hw99}
used {\it ROSAT} data to show that the soft X-ray emission had
relatively complex structure, with a small-scale extension around NGC
7237 as well as large-scale (3$'$, or 100 kpc) extension in a
direction perpendicular to the radio lobe axis.

\label{442-intro}

\section{Observations}

\subsection{X-ray observations}

3C\,285 and 3C\,442A were both observed using the ACIS instrument on
{\it Chandra}. 3C\,285 was observed for a total livetime of 39.3 ks in
a single observation with the ACIS-S in the HRC guaranteed
time program (PI: Stephen Murray), as part of a larger project to
complete {\it Chandra} observations of the 3CRR sample at $z<0.1$.
3C\,442A was observed in four separate observations, with three
distinct roll angles, with the ACIS-I as a guest-observer observation (PI:
Ralph Kraft) for a total livetime of 93.6 ks. Table \ref{obs} lists
the observing times and OBSIDs for the observations. We reprocessed
and filtered the data in the standard way, using the standard ASCA
grade set. Both observations were made in VFAINT mode and so we used
VF cleaning to reduce the background level. The 0.5-pixel event
position randomization in the standard pipeline processing was also
removed. We then used light curves excluding the nucleus and bright
point sources to search for intervals of high background. For 3C\,285
this flagged a small fraction of the data, as shown in Table
\ref{obs}.

\begin{deluxetable}{llllllll}
\tablecaption{Observations with {\it Chandra}}
\tablehead{Source&Detector&Chips on&Date&OBSID&Roll
  angle&Original&Filtered\\&&&&&(degree)&livetime (s)&livetime
  (s)}
\startdata
3C\,285&ACIS-S&23678&2006 Mar 18&6911&147.6&39625&39280\\[5pt]
3C\,442A&ACIS-I&01236&2005 Jul 27&5635&141.5&27006&27006\\
&&&2005 Jul 28&6353&141.5&13985&13985\\
&&&2005 Oct 07&6359&262.3&19884&19884\\
&&&2005 Dec 12&6392&312.2&32694&32694\\
\enddata
\label{obs}
\end{deluxetable}

\subsection{Radio data}

Both radio galaxies have been well observed with the Very Large Array
(VLA) in the radio at a number of frequencies \citep{vd93,co91} and
images showing the overall source structure at 1.4 GHz are available
from the online 3CRR atlas of J.P. Leahy
\etal\footnote{http://www.jb.man.ac.uk/atlas/}. We required
high-resolution radio images at multiple frequencies to allow us to
compare the radio data with the X-ray data at the highest resolution,
and we also needed multi-frequency data sampling the largest angular
scales of the sources to allow us to carry out internal pressure and
inverse-Compton calculations for the radio lobes. Accordingly, we
retrieved various VLA datasets from the online archive (see Table
\ref{vlaobs} for a complete list) and reduced them in the standard
manner within {\sc aips}. Where the VLA had observed the sources at a
given frequency at more than one configuration, we reduced the data
from each configuration separately, cross-calibrated for phase and
then concatenated the datasets with appropriate reweighting using the
{\sc aips} task DBCON. Mapping was carried out using the {\sc aips}
tasks IMAGR (for the lowest resolutions) and VTESS (for maps of
combined datasets at high resolutions). In addition, for flux
measurements of the lobes of 3C\,285 at low frequencies, we used Giant
Meter-wave Radio Telescope (GMRT) data at 240 and 610 MHz taken for
another purpose (PI: Martin Hardcastle). Maps from these observations
will be presented elsewhere.

\begin{deluxetable}{llllll}
\tablecaption{Observations with the VLA}
\tablehead{Source&Program ID&Date&Frequency&Configuration&Time on
  source\\
&&&(GHz)&&(h)
}
\startdata
3C\,285&AV127&1986 May 07&1.51&A&2.6\\
&AV127&1986 Aug 29&4.85&B&1.9\\
&AV127&1986 Aug 29&1.51&B&1.1\\
&AV127&1986 Dec 02&4.85&C&0.9\\
&AV127&1986 Dec 02&1.51&C&0.5\\
&AS549&1995 Apr 11&4.91&D&0.7\\
3C\,442A&AC131&1985 May 13&1.44&B&3.0\\
&AC131&1985 May 13&4.85&B&2.9\\
&AC131&1985 Jul 19&1.44&C&2.8\\
&AC131&1985 Jul 19&4.85&C&2.8\\
&AC131&1985 Dec 06&1.44&D&1.5\\
\enddata
\label{vlaobs}
\tablecomments{Observations were made at two observing frequencies.
  The frequency quoted is the mean of the two.}
\end{deluxetable}

\subsection{Optical and infrared data}

Both sources are close enough that images from the Digital Sky Survey
(DSS2) can be used to indicate the positions of the galaxies in the
host group. In addition, both have been observed with the {\it Hubble
Space Telescope} ({\it HST}), as discussed above, and with {\it
Spitzer} as part of a survey of the low-redshift 3CRR objects (PI:
Mark Birkinshaw). We use DSS2 and archival {\it HST} data, the latter
processed using the standard pipeline, where optical information is
important to our analysis. The {\it Spitzer} observations of 3C\,442A
are discussed by \citet{wkbh07} and we only refer to them in
passing in the present paper.

\section{Results}

In this section we discuss the main observational results from the
{\it Chandra} observations. We first briefly describe the small-scale
emission seen from the active nuclei of the systems, the detections of
or limits on any hotspot or jet-related components, and the other
galaxies in the group. We then discuss in detail the extended X-ray
emission seen in the {\it Chandra} data and its relationship to the
radio source.

All the X-ray analysis we describe was carried out using {\sc ciao}
3.3 and {\sc caldb} 3.2.2 for spectral extraction, and {\sc xspec}
11.3 for spectral fitting. The {\it specextract} script was used for
spectral extraction for extended sources, and the {\it psextract}
script for point sources. Galactic column densities of $1.37 \times
10^{20}$ cm$^{-2}$ and $5.08 \times 10^{20}$ cm$^{-2}$ are assumed for
3C\,285 and 3C\,442A respectively (interpolated from the data of
\citealt{sgwb92}) and are included in all spectral fits. We fitted to
data in the energy range 0.4--7.0 keV unless otherwise stated: spectra
were binned, typically to 20 net counts per bin after background
subtraction, so that $\chi^2$ statistics could be applied. Errors
quoted on parameters derived from {\sc xspec} fitting are the 90 per
cent confidence range for one interesting parameter, unless otherwise
stated, but errors quoted on measured numbers such as flux densities
or numbers of counts are the $1\sigma$ errors. Where counts are quoted
the $1\sigma$ errors are derived from the Gaussian ($\sqrt{n}$)
approximation to the true confidence range, as calculated by the {\sc
  funtools} code used to determine the background-subtracted values --
this is accurate to better than 10\% for $>10$ counts (cf.
\citealt{g86}), and the Gaussian assumption is necessary in any case
if we are to combine errors in the standard way with the uncertainty
on the expected background counts.

\subsection{Nuclei}

The X-ray nucleus of 3C\,285 has already been discussed by
\citet*{hec06}, using the same {\it Chandra} dataset. They found that
it was well modelled as a double power law, with one component being
heavily absorbed ($N_{\rm H} \approx 3 \times 10^{23}$ cm$^{-2}$) and
one having only Galactic absorption. The nucleus was relatively weak,
and so in their fits the poorly constrained power-law indices were
fixed to 2.0 and 1.7 for the unabsorbed and absorbed components
respectively. A combination of absorbed and unabsorbed emission is
consistent with what is seen for other narrow-line FRII radio
galaxies.

3C\,442A is a low-excitation radio galaxy \citep{lrl83} -- i.e., a
nuclear spectrum of NGC 7237 shows no strong high-excitation
emission-line features -- and so we might expect not to see a heavily
absorbed nuclear component \citep{hec06}. We extracted spectra for the
X-ray source corresponding to the currently active radio core using a
small (6 standard $0\farcs492$ {\it Chandra} pixels in radius)
circular extraction region and a concentric background annulus
(between 6 and 9 pixels). The nuclear spectra can be fitted acceptably
($\chi^2 = 12.6$ for 12 degrees of freedom) with a single power law
model with Galactic absorption and variable normalization (see below),
although the best-fitting single power-law photon index fitted to all
four spectra simultaneously is rather flat ($\Gamma = 0.9\pm 0.2$).
There is some evidence in radial profiles of the core region that even
this small-scale extraction region contains extended emission, and so
it is possible that the spectra are contaminated by some thermal
emission, but adding a thermal component does not significantly change
the best-fitting power-law spectrum. Adding some intrinsic absorption
(the same column density for all four epochs) gives a more typical
power-law index ($\Gamma = 1.4 \pm 0.5$) and a column density of
$(3.4_{-3.0}^{+4.5}) \times 10^{21}$ cm$^{-2}$, with a somewhat
improved fit ($\chi^2 = 8.6$ for 11 degrees of freedom). This column
density is high, but not unprecedentedly high, compared to other
low-power sources, and may indicate some obscuration by the
small-scale dust lane that is seen to cross the nucleus in the {\it
HST} observation. We estimate a maximum $A_V \sim 0.3$ for the off-nuclear
extinction seen in the {\it HST} data taken with the F547M filter,
which would be consistent with the lower end of the confidence range
for the X-ray absorbing column for a Galactic gas to dust ratio.

The core was clearly variable over the time period of our
observations, and acceptable fits were not obtained unless we allowed
the normalizations of the power laws fitted to each dataset to vary.
The normalizations for the four epochs listed in Table \ref{obs}
correspond to 1-keV flux densities of $3.2 \pm 0.8$, $2.9 \pm 0.9$,
$1.2 \pm 0.6$ and $4.5 \pm 0.7$ nJy respectively (assuming Galactic
absorption only: flux densities would be approximately a factor 2
higher for the best-fitting absorbing column). Variability of the
photon index as well as the normalization might explain the relatively
poor fit and flat photon index given by the combined dataset, but the
data are not good enough to constrain this possibility. We were
initially concerned that the variability might be an artefact of the
data, as the core in the third observation (obsid 6359) lies close to
a node boundary, with many events in the core region in the level 1
events file being flagged as being on or near a bad pixel. However, we
are satisfied that the ancillary response files (ARFs) make an
adequate correction for this reduced effective exposure time, and we
have verified that regions taken from the thermal emission close to
the core (\S\ref{tail-fit}) do not show any variability in count rate
or fitted parameters. We therefore infer that the core is genuinely
variable, presumably because of ongoing AGN activity. The 5-GHz flux
density of the core of 3C\,442A is 3.1 mJy in the VLA data we
use, compared to the value of 1.9 mJy from earlier radio observations
tabulated by \citet{fmtl84}, so that it appears that the radio core
may be similarly variable on long timescales (months in the
case of the X-ray, years in the case of the radio). The approximate
positions, given the uncertainties imposed by variability and
non-simultaneous observations, of the source on the radio/X-ray core
flux density and luminosity correlations for low-$z$ sources
\citep{ewhk06} are consistent with expectations. The bolometric
unabsorbed X-ray luminosity of the core region ranges between 2 and $7
\times 10^{41}$ erg s$^{-1}$, which puts it well above the
expectation from X-ray binaries from the whole galaxy \citep*{ofp01}
given the $B$-band optical luminosity of $\sim 3 \times 10^{10}
L_{{\rm B,}\odot}$ \citep{lrl83}.

\subsection{Jets, knots and hotspots}

\subsubsection{3C\,285}

\begin{figure}

\plotone{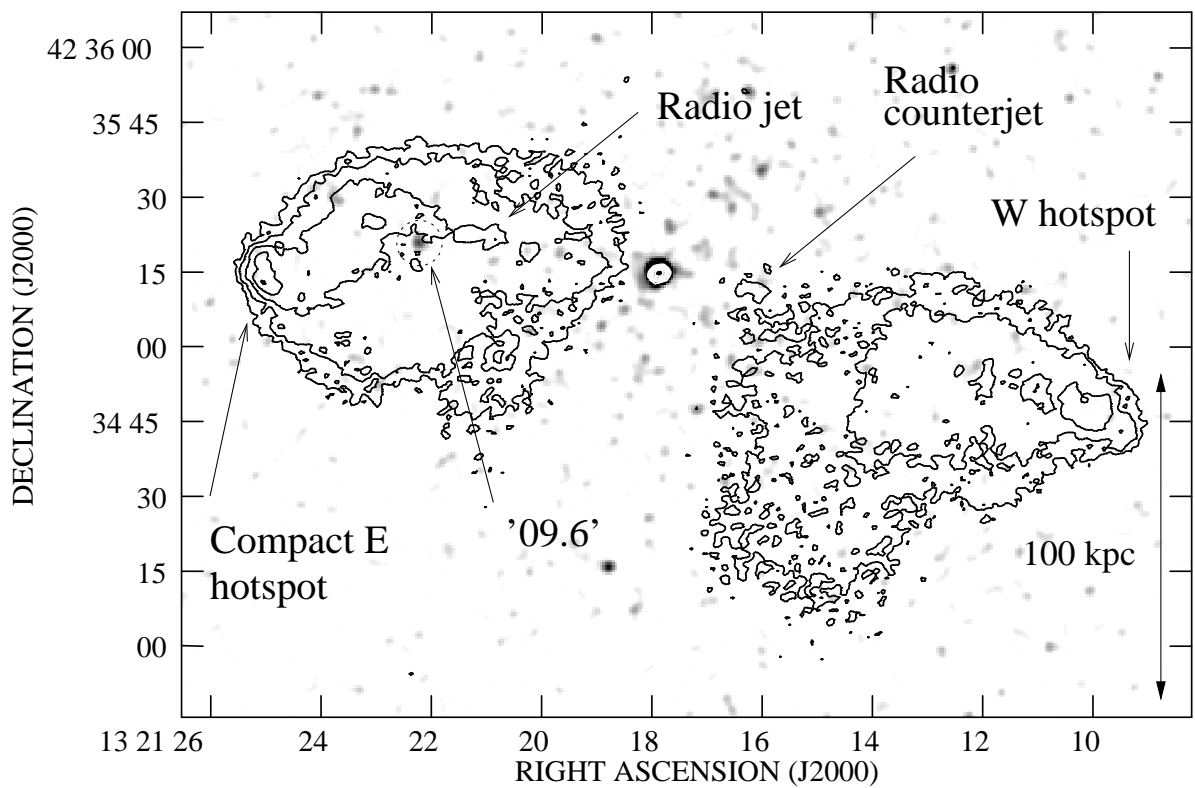}
\caption{Radio contours of 3C\,285 (4.9-GHz radio map with $1\farcs59
  \times 1\farcs29$ resolution) superposed on the 0.5--5.0 keV {\it
  Chandra} data smoothed with a $2\farcs0$ FWHM Gaussian. Compact features
  in the radio and X-ray are labeled. Only the radio core and the
  star-forming region 09.6 (see the text) are detected in the X-ray. A
  dotted circle marks the position of 09.6. Radio contours are at $0.1
  \times (1,2,4\dots)$ mJy beam$^{-1}$. Large-scale extended X-ray
  emission present in the data cannot be seen in this image.}
\label{285-highres-overlay}
\end{figure}

The jets and hotspots of some low-power FRII radio galaxies (e.g.,
3C\,403, \citealt{khwm05}) show prominent synchrotron X-ray
counterparts to compact radio features. 3C\,285 has a well-defined
though faint jet and counterjet in the radio and a bright compact
hotspot (in the E lobe). However, none of these is detectable as an
X-ray source in the Chandra observation (Fig.\
\ref{285-highres-overlay}). The only convincingly detected compact
feature in the lobes in the {\it Chandra} data (with $7 \pm 3$ counts
in the 0.5-5.0 keV energy band) is spatially coincident with the blue
object denoted 09.6 by \citet{vd93}, lying within $1''$ of the peak of
the resolved optical emission in the {\it HST} images, with a
centroiding uncertainty of $0\farcs5$. 09.6 was identified by
\citet{vd93} as a star-forming galaxy that is at the redshift of
3C\,285's host, whose star formation may be related to the nearby
radio jet. Roughly, the unabsorbed bolometric X-ray luminosity of this
object in the {\it Chandra} band is of the order of $10^{40}$ erg
s$^{-1}$ (assuming a MEKAL model with $kT = 0.5$ keV), which is a
plausible luminosity for a dwarf starburst system with $M_B \approx
-18.5$ (the absolute magnitude quoted by \citeauthor{vd93}: our
different cosmology makes no significant difference): for example, it
is comparable to, though somewhat higher than, the luminosities
derived for a sample of nearby, generally somewhat optically fainter
dwarf starbursts by \citet*{owb05}. The X-ray detection is thus
consistent with the starburst interpretation for 09.6.

\begin{figure}
\plottwo{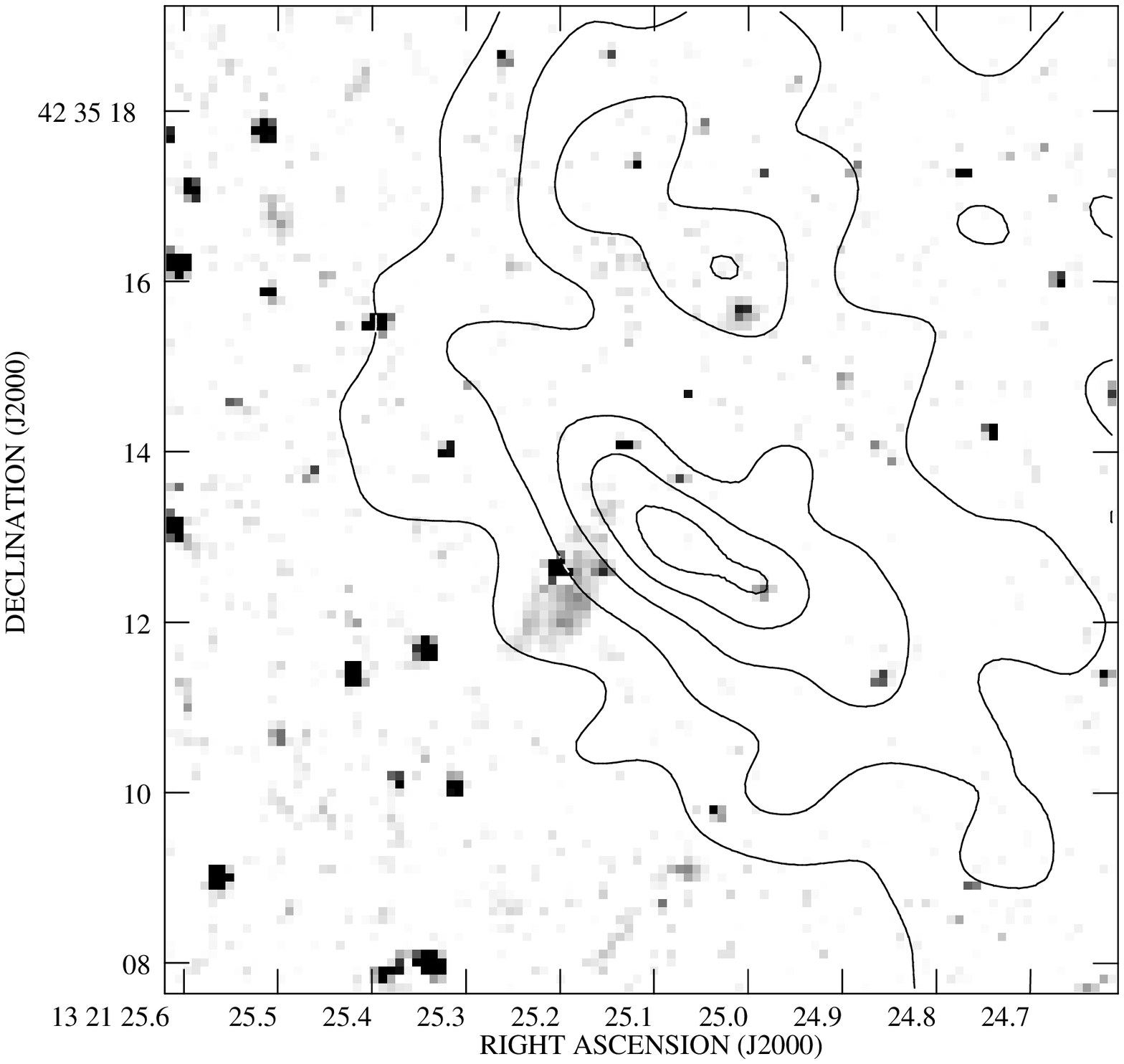}{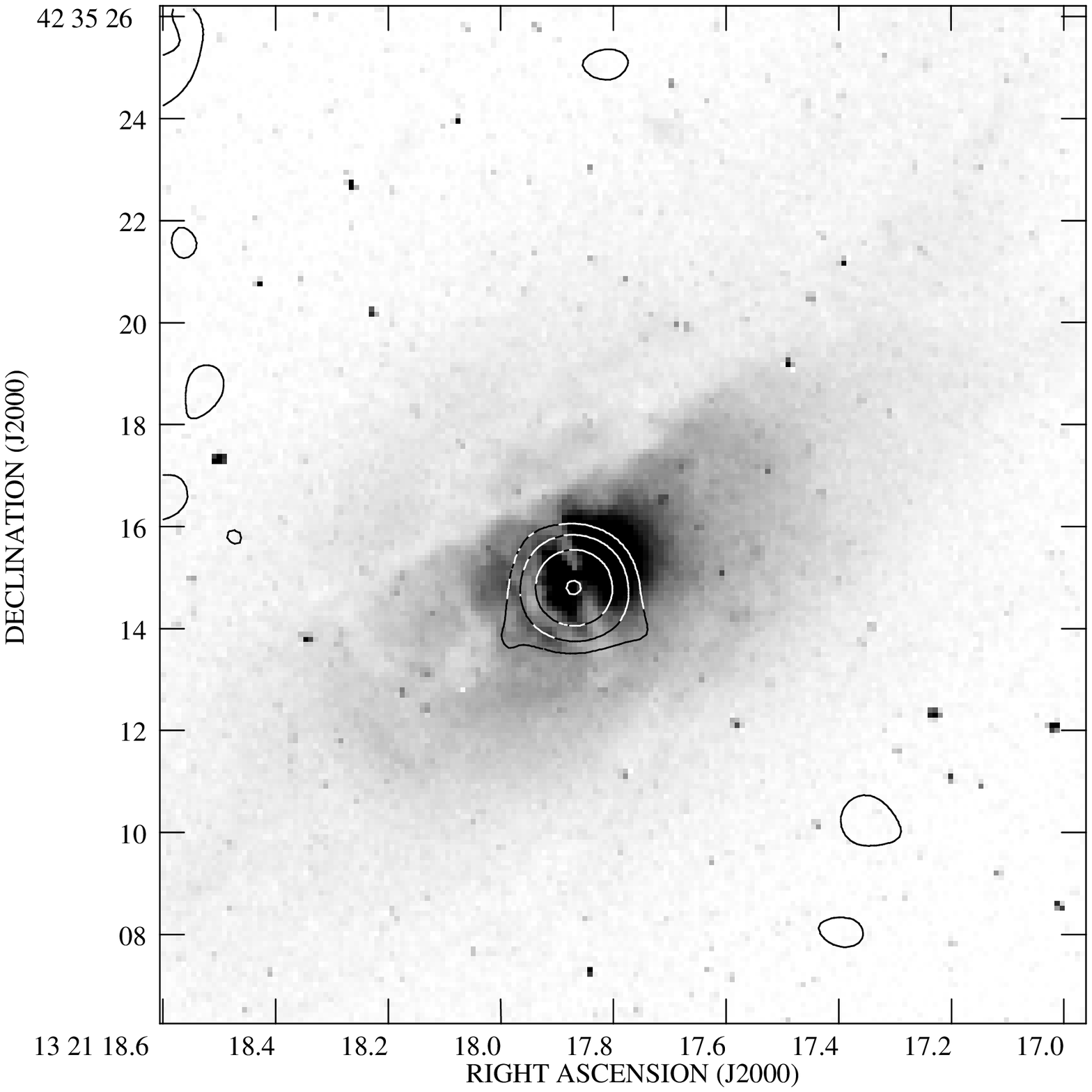}
\caption{Two {\it HST} images of small regions of 3C\,285 with superposed radio
  contours. The {\it HST} data is the F702W snapshot survey
  observation, the radio map is made from the 4.9-GHz data with a
  $1''$ circular restoring beam. On the right, the good alignment between the
  radio core and the center of the host galaxy can be seen: the
  complex dusty structure of the host galaxy is also apparent. On the
  left is the bright E hotspot: the nearest optical counterpart is
  clearly offset from, and morphologically dissimilar to, the radio
  emission. Contours of the radio emission are at (left) $0.2 \times
  (1,2,3,\dots)$ mJy beam$^{-1}$ and (right) $0.1 \times
  (1,4,16\dots)$ mJy beam$^{-1}$. The high density of cosmic rays to
  the left of the left-hand image is due to the fact that part of the
  relevant WF chip was not exposed in one of the two observations that
  make up the {\it HST} data: this does not affect the validity of
  detections of extended structure.}
\label{285-hst}
\end{figure}

The non-detections of X-ray emission from the jets and hotspots in
3C\,285 correspond to upper limits on the X-ray flux density. Roughly,
a $3\sigma$ upper limit on the total background-subtracted counts in
the 0.5-5.0 keV energy range from a region corresponding to the
brightest compact hotspot (in the E lobe) is $<4$ counts: this
corresponds to a flux density limit of $<0.1$ nJy at 1 keV for a
source with photon index 2.0. However, the radio flux densities for
the hotspot components are faint compared to the sources in which
X-ray emission has been seen. The total 5-GHz flux density of the E
hotspot is 6 mJy, which implies that $\alpha_{RX} > 1.01$ (where
$\alpha_{RX}$ is the two-point radio-to-X-ray energy spectral index:
$\alpha$ is defined here and throughout the paper in the sense that
flux density $\propto \nu^{-\alpha}$), but this is not a very strong
constraint, as there are some hotspots with detected synchrotron X-ray
emission that show $\alpha_{\rm RX} > 1.0$ \citep{hhwb04}, while we
would expect any synchrotron self-Compton emission at equipartition to
be at the level of about 10 pJy. The E hotspot of 3C\,285 is
potentially interesting as it is the one claimed by \citet{stc78} to
show an optical counterpart. However, examination of the F702W {\it
HST} observations (originally discussed by \citealt{mbsw99}) shows
that the peak of the optical emission from the hotspot region is
offset by $\sim 2''$ from the peak of the radio emission, using the
default astrometry, which gives a reasonably good alignment (offset
$\sim 0\farcs3$) between the radio core and the center of the galaxy
(Fig.\ \ref{285-hst}). More importantly the optical object is
morphologically quite different from the radio structure, and seems
most likely to be a faint irregular galaxy. We thus conclude that
there is no evidence for any emission from the E hotspot of 3C\,285 in
any band other than the radio. The W hotspot is fainter in the radio,
so again we can only put weak limits on the radio to X-ray spectral
index, and it has no detectable optical counterparts. Overall, the
hotspots of 3C\,285 seem to present a rather different picture to
those seen in other nearby FRII radio galaxies of similar power such
as 3C\,403 \citep{khwm05}, 3C\,321 \citep{hhwb04} and 3C\,33
\citep{kbhe07}. The hotspots of these objects, which have radio
luminosities spanning a range around that of 3C\,285, often have
optical/IR and X-ray counterparts; their X-ray emission, largely
inferred to be of synchrotron origin, has $0.95 < \alpha_{RX} < 1.1$.
The lack of an optical counterpart in particular may suggest that
3C\,285's hotspots are not able to accelerate particles to even
moderately high energies. However, the biased nature of existing
observations of nearby radio galaxies, many of which have been
selected for observation because of their prominent radio or optical
hotspots, makes it hard to say which is more typical.
\label{opt-hotspot}

\subsubsection{3C\,442A}

\begin{figure}

\plotone{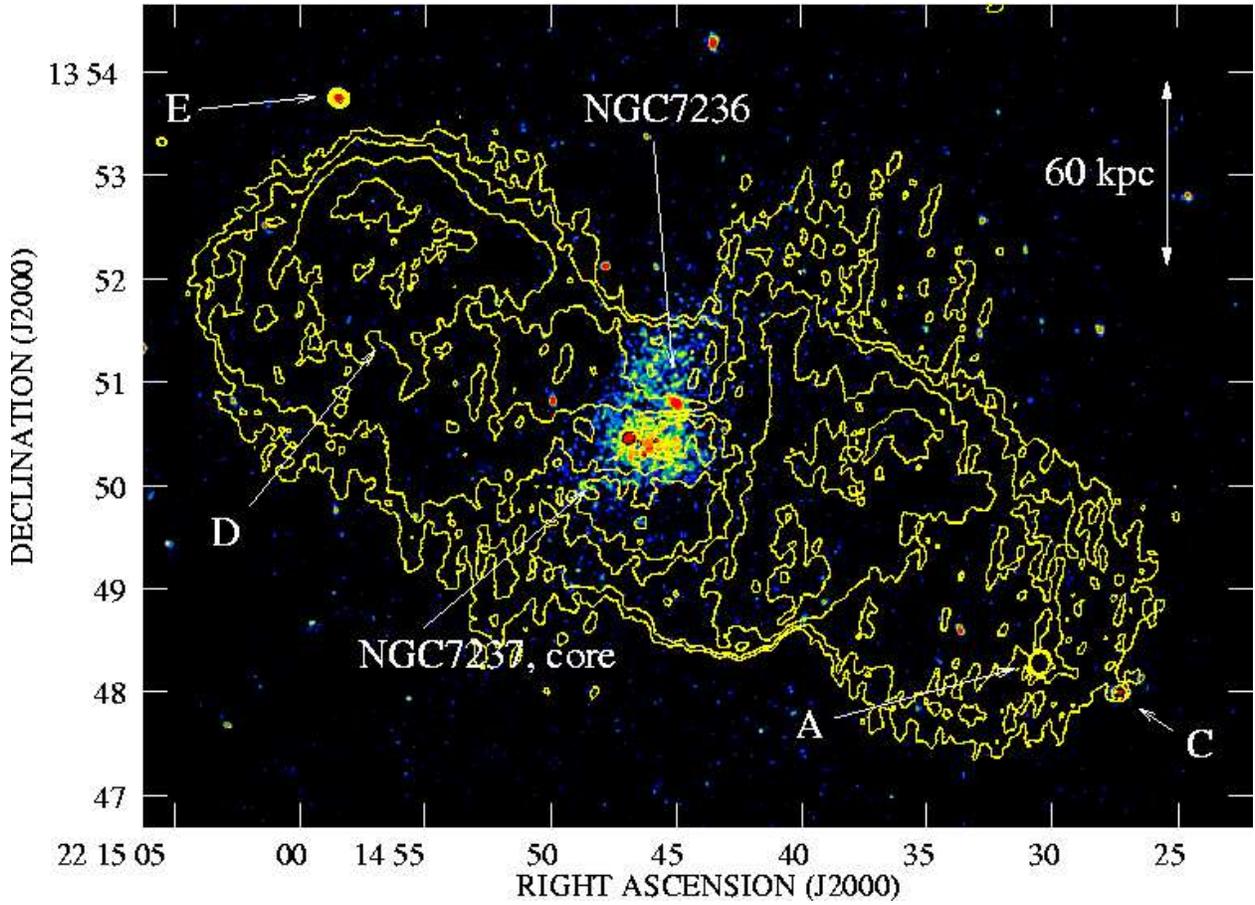}
\caption{Radio contours of 3C\,442A (1.4-GHz radio map with $5\farcs37
  \times 4\farcs35$ resolution) superposed on the 0.5--5.0 keV {\it
  Chandra} data from all four observations smoothed with a $2\farcs0$
  Gaussian. Compact features in the radio and X-ray are labeled.
  Large-scale extended X-ray emission present in the data cannot be
  seen in this image.}
\label{442-highres-overlay}
\end{figure}

As discussed in \S\ref{442-intro}, it is not clear whether any of the
bright compact radio sources in the 3C\,442A lobe are related to the radio
galaxy itself. The X-ray data shed some light on this question, but do
not completely resolve it. Fig.\ \ref{442-highres-overlay} shows our
highest-resolution radio map overlaid on lightly smoothed {\it
Chandra} data in the 0.5-5.0 keV band, with point sources labeled
using the notation of \citet{co91}. It can be seen that the two point
sources, C and E, that lie outside or just on the edge of the lobes
are relatively strong X-ray sources, having respectively $215 \pm 16$
and $176 \pm 16$ net counts in total in our observations. As
\citeauthor{co91} report, both these sources have optical
counterparts, and it seems most likely that they are background AGN of
some kind. C has an inverted radio spectrum between 1.4 and 4.9 GHz
($\alpha = -0.3$) while E is steep-spectrum ($\alpha = 1.1$). The most
puzzling of the bright point sources, A, is not detected in the X-ray
and has no optical counterpart on DSS2 images (neither the {\it HST}
nor {\it Spitzer} fields of view are large enough to give us more
sensitive infrared or optical constraints). It is resolved at the highest radio
resolution we have available ($1\farcs4$) into two components
separated by $2\farcs6$, the northern one unresolved and containing
most of the flux, the southern one slightly resolved. Its radio
spectrum is steep ($\alpha = 1.2$), consistent with the results of
\citet{blp81} at lower frequencies, which might imply that it is part
of the source: but the lack of any connection to the filamentary
structure in the lobes remains a strong argument against this, and it
might equally well be a high-redshift double or core-jet source. If the
source were a hotspot, the $3\sigma$ upper limit of 6 net counts in
our observations, corresponding to 0.1 nJy at 1 keV for a source with
photon index 2.0, would imply $\alpha_{\rm RX} > 1.03$, but again this
is not incompatible with the behaviour of some hotspots. No X-ray
counterpart is detected for the weak point source D discussed by
\citeauthor{co91}, but it seems likely to be part of the complex
filamentary structure in that part of the lobe, rather than a true
point source. We do not detect the source B, referred to by
\citeauthor{co91}, in our radio images.

\subsection{Group members}

\subsubsection{3C\,285}

We used the NASA Extragalactic Database,
NED\footnote{http://nedwww.ipac.caltech.edu/}, to search for optical
objects identified as galaxies around the position of 3C\,285 and,
where possible, to obtain their redshifts as determined by the Sloan
Digital Sky Survey\footnote{http://www.sdss.org/}. In a $5'$ search
radius (chosen to cover the ACIS-S3 chip: equivalent to 450 kpc at the
distance of 3C\,285) there are 13 objects classified as galaxies
(including 3C\,285 itself) of which 8 have measured radial velocities:
6 of these are possibly consistent with membership of the 3C\,285
group, while the other two are foreground objects. The positions and
redshifts of these nearby galaxies are listed in Table \ref{285-sdss}.
A few of these galaxies are detected in X-rays in the {\it Chandra}
data. We tabulate counts measured from the events file in $10''$ (15
kpc) extraction regions with background subtraction from a larger
local background region. Upper limits are $3\sigma$ values based on
Poisson statistics and derived from the local background value.

\begin{deluxetable}{lllll}
\tablecaption{Galaxies in the 3C\,285 group}
\tablehead{Galaxy name&RA&Dec.&Radial velocity&X-ray counts\\&&&(km
  s$^{-1}$)&(0.5-5.0 keV)}
\startdata
3C\,285 &$13\hh21\mm17\fs8$ & $+42\degr35'15''$ &  23804 & $273 \pm 16$ \\
SDSS J132116.31+423551.3 &$13\hh21\mm16\fs3$ & $+42\degr35'52''$ &  --
  & $22 \pm 6$\\
SDSS J132114.49+423620.3 &$13\hh21\mm14\fs5$ & $+42\degr36'20''$ & --
  & $15 \pm 5$\\
SDSS J132114.01+423557.5 &$13\hh21\mm14\fs0$ & $+42\degr35'58''$ &
  23738 & $<11$\\
SDSS J132114.75+423631.1 &$13\hh21\mm14\fs7$ & $+42\degr36'31''$ &
  24623 & $12 \pm 5$ \\
SDSS J132117.98+423657.4 &$13\hh21\mm18\fs0$ & $+42\degr36'57''$ &
  24130 & $14 \pm 5$\\
SDSS J132130.21+423544.3 &$13\hh21\mm30\fs2$ & $+42\degr35'44''$ &
  22643 & $<11$ \\
MAPS-NGP O-219-0156129 &$13\hh21\mm03\fs3$ & $+42\degr36'47''$ &  -- & $<8^*$ \\
MAPS-NGP O-219-0183438 &$13\hh21\mm38\fs7$ & $+42\degr34'19''$ &  -- &$<11$\\
SDSS J132131.68+423208.2 &$13\hh21\mm31\fs7$ & $+42\degr32'08''$ &  10678 &$<11$\\
MAPS-NGP O-219-0156725 &$13\hh21\mm31\fs9$ & $+42\degr39'08''$ &  -- &
  -- \\
SDSS J132051.31+423550.5&$13\hh20\mm51\fs3$ & $+42\degr35'50''$ &
  15351 & $13 \pm 5^*$ \\
SDSS J132058.74+423147.6 &$13\hh20\mm58\fs7$ & $+42\degr31'48''$ &
  22115 & -- \\
\enddata
\tablecomments{Galaxies are sorted by distance from 3C\,285. Galaxies with no {\it Chandra} detection or upper limit
  lie partly or wholly off the detectors. Count measurements or upper
  limits marked with an asterisk are taken from the front-illuminated
  ACIS-S4 chip. Where objects are named by their 2MASS name in NED we
  have used the SDSS name instead to make the origin of the data
  clearer. SDSS J132114.49+423620.3 is not included in the NED listing
but is clearly a nearby member of the group, so we have taken its name
and position directly from the SDSS DR5. All counts tabulated are
background subtracted: upper limits are $3\sigma$.}
\label{285-sdss}
\end{deluxetable}

Although the two bright galaxies adjacent to 3C\,285 have no
spectroscopic information available in SDSS, we can estimate the
velocity dispersion of the group from the information we have. If we
include all SDSS velocities with $22000 < cz < 25000$ km s$^{-1}$,
then the mean heliocentric radial velocity is $23500$ km $s^{-1}$, and
the velocity dispersion $\sigma$ is $900_{-200}^{+500}$ km s$^{-1}$,
which would be high for a group. However, as Table \ref{285-sdss}
shows, the two galaxies furthest away from 3C\,285 have radial
velocities that differ markedly from its velocity. If we consider only
the four galaxies closest to 3C\,285 on the sky then the mean velocity
is 24070 km s$^{-1}$, about 270 km s$^{-1}$ higher than that of
3C\,285 itself, but the velocity dispersion is a more reasonable
$400^{+400}_{-100}$ km s$^{-1}$ (where the errors are determined using
the method of \citealt*{ddd80}). We cannot draw many conclusions from
such a small number of radial velocities (a relatively short optical
observation would give us much better constraints on the nature of the
optical group) but it seems likely that only the inner galaxies in
this region should be considered to be bound group members. Using the
relation of \citet{hp00} the lower velocity dispersion would be
expected to correspond to a temperature for the X-ray gas of
$1.1_{-0.3}^{+0.9}$ keV, where the errors here are only statistical
and take no account of the scatter in the $\sigma$ -- $T$ relation.
\label{285-vd}

\subsubsection{3C\,442A}

As for 3C\,285, we used NED to search for galaxies in the 3C442A
group. As 3C\,442A is substantially closer to us, the data are better:
in a $14'$ (450 kpc) radius aperture there are 40 galaxies, all of
which have measured radial velocities from the SDSS or elsewhere. 31
of these have radial velocities within 1000 km s$^{-1}$ of the value
for NGC7236/7, and so are potential group members. Positions and
radial velocities of these objects are tabulated in Table
\ref{442a-sdss}.

\begin{deluxetable}{lllll}
\tablecaption{Galaxies in the 3C\,442A group}
\tablehead{Galaxy name&RA&Dec.&Radial velocity&X-ray counts\\&&&(km
  s$^{-1}$)&(0.5-5.0 keV)}
\startdata
NGC 7237 &$22\hh14\mm46\fs8$ & $+13\degr50'27''$ &  7868 & $483 \pm  31$\\
NGC 7236 &$22\hh14\mm45\fs0$ & $+13\degr50'47''$ &  7879 &$222 \pm  23$\\
NGC 7237C &$22\hh14\mm48\fs8$ & $+13\degr50'01''$ &  7153 &$ 37 \pm  13$\\
SDSS J221439.24+134920.0 &$22\hh14\mm39\fs2$ & $+13\degr49'20''$ &  7997&$<21$ \\
SDSS J221437.58+135053.0 &$22\hh14\mm37\fs6$ & $+13\degr50'53''$ &  8375&$<19$ \\
2MASX J22144047+1352416 &$22\hh14\mm40\fs4$ & $+13\degr52'42''$ &  8030&$<20$ \\
SDSS J221437.45+135228.9&$22\hh14\mm37\fs4$ & $+13\degr52'29''$ &  8280&$<19$\\
SDSS J221432.20+134825.4 &$22\hh14\mm32\fs2$ & $+13\degr48'25''$ &  7679&$<20$ \\
SDSS J221437.45+135229.0&$22\hh14\mm39\fs5$ & $+13\degr46'27''$ &  7915&$<19$ \\
SDSS J221451.49+135543.8 &$22\hh14\mm51\fs5$ & $+13\degr55'44''$ &  35633 &$<15$\\
CGCG 428-060 &$22\hh14\mm57\fs8$ & $+13\degr45'43''$ &  7270 & $ 34 \pm  10$\\
LSBC F673-02 &$22\hh14\mm31\fs9$ & $+13\degr46'22''$ &  7293 & $ 33 \pm  10$\\
SDSS J221445.84+134449.8&$22\hh14\mm45\fs8$ & $+13\degr44'50''$ &  8161 &$<19$\\
SDSS J221509.33+134813.3 &$22\hh15\mm09\fs3$ & $+13\degr48'13''$ &  7952&$<18$ \\
SDSS J221424.81+134633.4 &$22\hh14\mm24\fs8$ & $+13\degr46'33''$ &  7263 &$<24$\\
SDSS J221445.10+134304.9&$22\hh14\mm45\fs1$ & $+13\degr43'05''$ &  8207 &$<16$\\
CGCG 428-059 &$22\hh14\mm51\fs4$ & $+13\degr42'55''$ &  7245 & $ 20 \pm  12$\\
SDSS J221514.93+135419.6 &$22\hh15\mm14\fs9$ & $+13\degr54'20''$ &  17810 &$<27$ \\
SDSS J221413.65+135044.0 &$22\hh14\mm13\fs6$ & $+13\degr50'44''$ &  120542&$<21$ \\
SDSS J221444.51+135836.6 &$22\hh14\mm44\fs5$ & $+13\degr58'37''$ &  7920&$<20$ \\
LCSB S2633P &$22\hh14\mm14\fs2$ & $+13\degr52'30''$ &  7222 &$<20$\\
SDSS J221414.03+135232.4&$22\hh14\mm14\fs0$ & $+13\degr52'32''$ &  7276&$<21$ \\
SDSS J221435.72+135822.6&$22\hh14\mm35\fs7$ & $+13\degr58'23''$ &  8297&$<18$ \\
CGCG 428-054 &$22\hh14\mm21\fs8$ & $+13\degr57'11''$ &  7795 &$37 \pm  11$\\
SDSS J221508.74+135815.1 &$22\hh15\mm08\fs7$ & $+13\degr58'15''$ &  7799 &$ 99 \pm  22$\\
SDSS J221450.80+134051.3 &$22\hh14\mm50\fs8$ & $+13\degr40'51''$ &  7374 &$<25$\\
SDSS J221422.18+134248.3 &$22\hh14\mm22\fs2$ & $+13\degr42'48''$ &  18124&$ 25 \pm  12$ \\
SDSS J221527.93+135114.3&$22\hh15\mm27\fs9$ & $+13\degr51'14''$ &  6634 &$<29$\\
SDSS J221530.83+134912.5 &$22\hh15\mm30\fs8$ & $+13\degr49'13''$ &  115271&$<21$ \\
SDSS J221404.42+134601.3 &$22\hh14\mm04\fs4$ & $+13\degr46'01''$ &  7645 &$<31$\\
SDSS J221414.10+135901.2 &$22\hh14\mm14\fs1$ & $+13\degr59'01''$ &  29751&$ 84 \pm  23$ \\
SDSS J221514.34+134045.3 &$22\hh15\mm14\fs3$ & $+13\degr40'45''$ &  8356 &--\\
SDSS J221407.71+134230.7 &$22\hh14\mm07\fs7$ & $+13\degr42'31''$ &  7361&-- \\
SDSS J221402.08+134424.9 &$22\hh14\mm02\fs1$ & $+13\degr44'25''$ &  120609&-- \\
SDSS J221515.54+140059.1 &$22\hh15\mm15\fs5$ & $+14\degr00'59''$ &  7956&-- \\
SDSS J221540.36+135255.5&$22\hh15\mm40\fs3$ & $+13\degr52'56''$ &  7434&-- \\
SDSS J221509.85+140230.4&$22\hh15\mm09\fs8$ & $+14\degr02'30''$ &  8314&$<71$ \\
SDSS J221356.95+135601.3 &$22\hh13\mm56\fs9$ & $+13\degr56'01''$ &  31005&$<37$ \\
UGC 11953 &$22\hh13\mm55\fs9$ & $+13\degr45'19''$ &  7585&--\\
SDSS J221542.30+134855.7&$22\hh15\mm42\fs3$ & $+13\degr48'56''$ &  7923&$<37$ \\
\enddata
\tablecomments{Galaxies are sorted by distance from NGC 7237. Galaxies
  with no {\it Chandra} detection or upper limit lie partly or wholly
  off the detectors. Where objects are named by their 2MASS name in
  NED we have used the SDSS name instead to make the origin of the
  data clearer.}
\label{442a-sdss}
\end{deluxetable}

Several of the other galaxies in the 3C\,442A group, and one or two of
the background galaxies, are detected in the {\it Chandra} image. As
with 3C\,285, we measure counts in a $10''$ aperture with a larger
local background region, and these are tabulated in Table
\ref{442a-sdss}. Because the response is non-uniform over the field
covered by the observations, as a result of the multiple {\it Chandra}
pointings, we have used the exposure maps to correct all the counts
tabulated to be equivalent to those measured at NGC 7237 with the full
exposure of the combined dataset, by scaling the measured counts and
error up by the ratio of the exposure map value at the position of the
source to the exposure map value at the position of NGC 7237. Where
there is no detection, an upper limit is tabulated, which is the
$3\sigma$ value determined using Poisson statistics based on the local
background count density and then scaled by the exposure in the same
way: upper limits are thus higher for regions with shorter exposure. In all
cases background-subtracted, scaled counts or limits are rounded to
the nearest integer.

The only galaxy other than NGC 7237 bright enough for spectroscopy is
NGC 7236. The nucleus of this source is a compact X-ray source with
$89 \pm 13$ net counts in the 0.5--5.0 keV band (using a $3''$ source
circle and adjacent background). Because these counts are spread over
the four observations, we had to bin the spectra to 10 net counts per
bin (rather than our usual 20) to give us enough bins for spectral
fitting. We then fitted the joint spectrum derived from the four
datasets. The spectrum is equally well fitted with a very steep power
law or with a thermal (MEKAL) model with $kT = 0.5\pm 0.1$ keV, both
with Galactic absorption, the latter giving $\chi^2 = 5.6$ for 6
d.o.f. with abundance fixed to 0.5 solar. The unabsorbed bolometric
X-ray luminosity in the thermal model was $(2 \pm 1) \times 10^{40}$
erg s$^{-1}$. There was no evidence for any variation of the
normalization of the spectrum between the datasets. Thus there is no
evidence for any AGN-related emission in the center of NGC 7236.

Using the subset of 31 galaxies with $6800<cz<8800$ km s$^{-1}$ we
determine the mean heliocentric radial velocity of the group to be
7768 km s$^{-1}$, within 100 km s$^{-1}$ of the velocity of NGC 7236/7
itself. The velocity dispersion is $390^{+60}_{-40}$ km s$^{-1}$, so
we would expect a temperature for the intragroup gas of $1.0\pm0.1$
keV using the $\sigma$--$T$ relation of \citet{hp00}.
\label{442-vd}

Both SDSS quasars in the field (SDSS J221458.45+135344.7, with
$z=3.67$, and SDSS J221453.84+140022.2, $z=1.52$) are clearly detected
in X-rays, with $197 \pm 16$ and $430 \pm 30$ exposure-corrected
0.5-5.0 keV counts respectively.

\subsection{Extended emission}
\label{group-properties}

\subsubsection{3C\,285}

\begin{figure}
\plotone{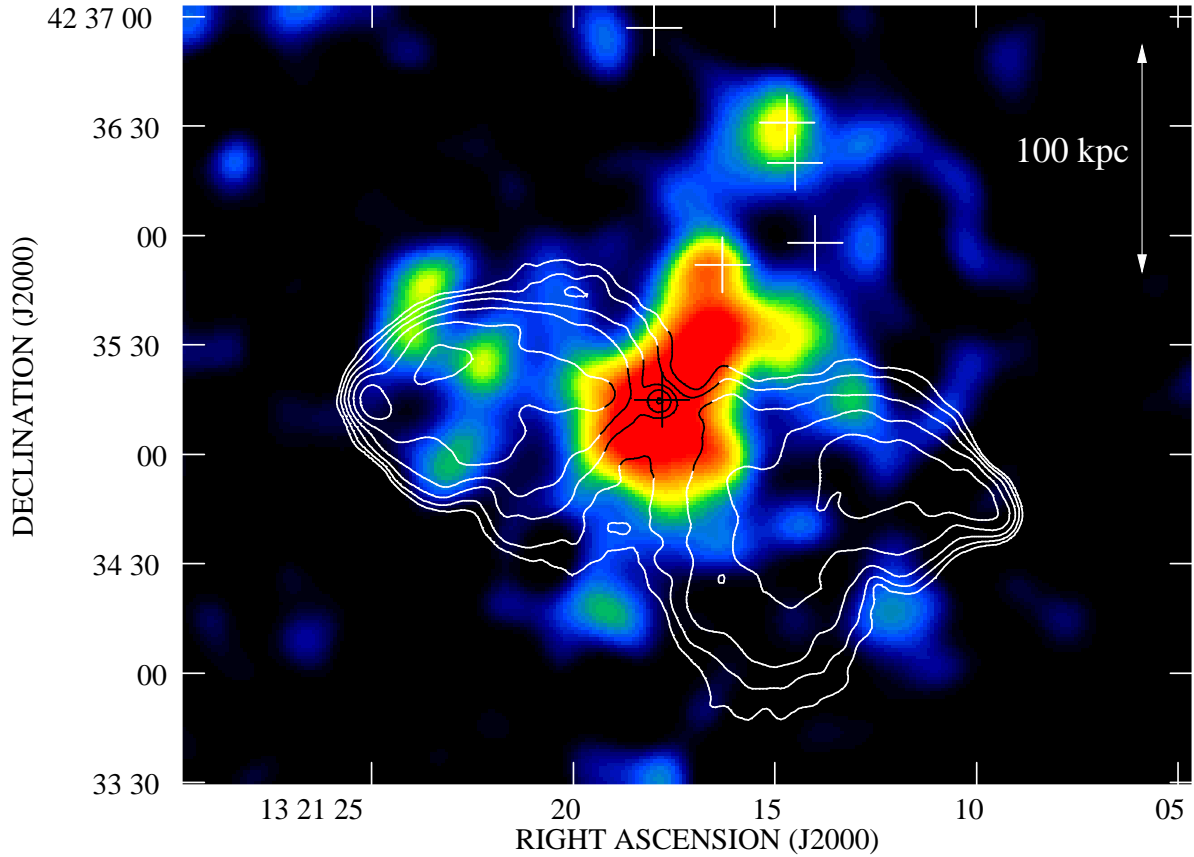}
\caption{Extended emission in the 3C\,285 group. The colors show the
  {\it Chandra} data after removal of compact objects, including the
  detected galaxies and the nuclear point source, and smoothing with a
  FWHM$=15''$ Gaussian. Overlaid are contours from an 1.5-GHz map at
  $5''$ resolution at $0.5 \times (1,2,4\dots)$ mJy beam$^{-1}$.
  Crosses show the positions of nearby galaxies in the group taken
  from Table \ref{285-sdss}}
\label{285-smooth}
\end{figure}

Clear asymmetrical extended X-ray emission is visible on arcminute
scales around 3C\,285 (Fig.\ \ref{285-smooth}). Near the nucleus, the
elongation is in a similar direction to the elongation of the optical
axis of the host galaxy: on larger scales, it extends in the direction
of the other nearby galaxies in the group, and essentially
perpendicular to the radio source axis. The central region of
emission, roughly corresponding to the red-colored region in (Fig.\
\ref{285-smooth}, extends for just over $1'$ (90 kpc in projection)
but the X-ray emission to the NW and SE of the main extended region is
connected to it by other X-ray emission at $\sim 99$ per cent
confidence, so it is possible that we should view the whole region
from SE to NW of the nucleus as forming one physical bar or disk of
X-ray-emitting material with varying density, with a size scale of up
to $2\farcm5$ (220 kpc). For convenience, and without presupposing any
particular physical interpretation, we follow \citet{opf04} and refer
to the structure as the `ridge' in what follows.

We extracted a spectrum for the ridge from a $1\farcm7 \times
0\farcm8$ rectangle centred on the host galaxy, with the long axis in
position angle $334^\circ$ (measured N through E). This region
contains $245 \pm 22$ net 0.5--5.0 keV counts, using off-source
background, and we found that it was acceptably fitted ($\chi^2 =
11.7$ for 10 d.o.f.) with a single-temperature MEKAL model with fixed
abundance of 0.35 solar (we adopt this value as the data are not
adequate to constrain abundance) and $kT = 1.07_{-0.11}^{+0.24}$ keV.
The bolometric unabsorbed X-ray luminosity of the ridge on this model
is $(5.5 \pm 0.5) \times 10^{41}$ erg s$^{-1}$. The residuals for this
model show a soft excess which can be modeled as a $\sim 0.2$-keV
thermal plasma, also with 0.35 abundance (the temperature is not well
constrained, but $kT < 0.35$ keV at 90 per cent confidence) without
affecting the temperature of the hot component, $kT =
1.08_{-0.13}^{+0.27}$ keV: this gives an improvement in the fit to
$\chi^2 = 4.8$ for 8 d.o.f.. A somewhat less good fit is obtained
($\chi^2 = 8.5$ for 9 d.o.f.) if the second component in the
two-temperature model is fixed to the temperature (0.64 keV) derived
from fits to the group-scale gas (see below): in this case the hotter
component has $kT = 1.3_{-0.3}^{+0.6}$ keV. The hotter component is
always the dominant one in these fits, and the addition of a second
thermal component reduces the emission measure by at most 10\%.

Some weak excess emission is also spatially coincident with the lobes
of 3C\,285 (this is just visible in parts of the lobe in Fig.\
\ref{285-smooth}). We extract spectra in two circular regions matched
to the E and W lobes, excluding point sources and the region of the
ridge described above, and taking a background annulus concentric with
the lobes. We find $107 \pm 23$ counts in the E lobe and $31 \pm 28$
counts in the W lobe: thus only the E lobe is significantly detected.
Fitting a power-law model to the E lobe's spectrum we find a photon
index of $1.6 \pm 0.5$ ($1\sigma$ error: $\chi^2 = 1.3$ for 2 d.o.f.)
which is consistent with the expectations from the inverse-Compton
process, although of course the errors are large and the spectrum
would be equally consistent with a hot thermal model. Assuming the
power-law model and best-fitting photon index, the 1-keV flux density
of the E lobe is $3 \pm 1$ nJy, and, scaling by count rate, $1 \pm 1$
nJy for the W lobe. The predicted 1-keV flux density of emission from
inverse-Compton scattering of the cosmic microwave background
radiation (CMBR) for equipartition between magnetic fields and
radiating electrons (using the code of \citealt*{hbw98}) is 0.6 nJy for
the E lobe and 0.9 nJy for the W lobe: thus, if all the emission from
the E lobe is inverse-Compton in origin, the lobe must be somewhat out
of equipartition, though not by a particularly unusual amount in the
context of FRII sources in general \citep{chhb05b}.

\begin{figure}
\plotone{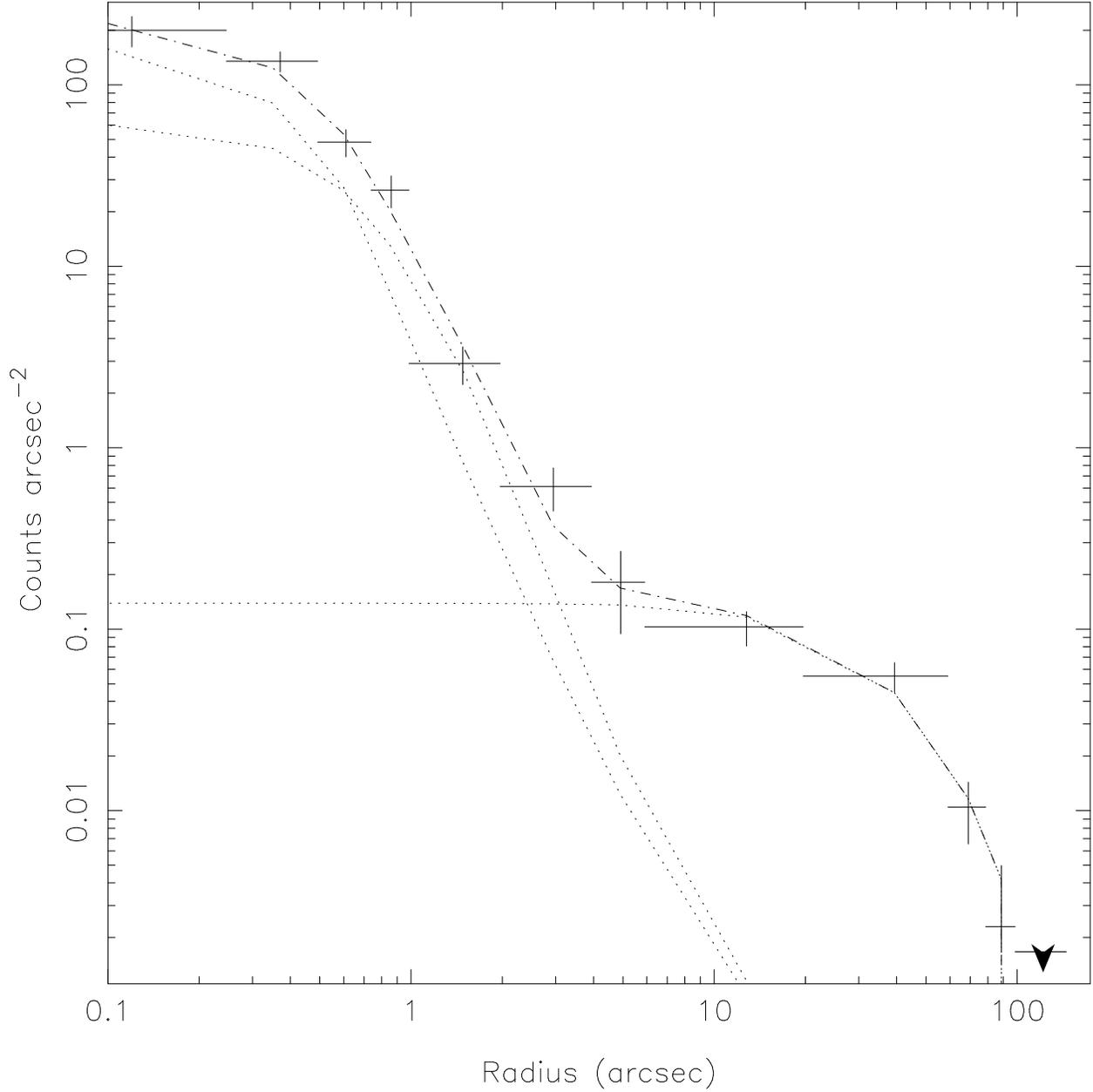}
\caption{Background-subtracted radial profile of extended emission in
  the inner 100'' of 3C\,285 after masking of lobes and asymmetrical
  structure. The dotted lines show the three components of the
  best-fitting model described in the text, and the dash-dotted line
  their sum.}
\label{285-profile}
\end{figure}

On larger scales, there is some excess emission on $100''$ (150-kpc)
scales even when the `ridge', the lobes and all the point sources
(including all those visible on Fig.\ \ref{285-highres-overlay}) are
masked out. In a $100''$ source circle with an adjacent background
annulus (the largest circular region that can be examined without
going off the S3 chip) there are $100 \pm 34$ net 0.5--5.0 keV counts.
Fitting a spectrum to this region, we found it to be adequately
modeled ($\chi^2 = 5.4$ for 5 d.o.f.) with a MEKAL model, again with
fixed abundance of 0.35 solar, with $kT = 0.64 \pm 0.25$ keV. This
temperature is consistent with the rough prediction of the group gas
temperature made in \S\ref{285-vd}. We used radial profiling (again
masking out the lobes and elongated central region, but not masking
the central few arcsec to allow fitting to the central point source),
correcting for the small changes in {\it Chandra}'s response over this
radial distance with an exposure map, to determine rough structural
parameters of this extended component, fitting $\beta$ models
convolved with the {\it Chandra} PSF in the manner described by
\citet{bw93} (using the PSF models of
\citealt{wbh01}) to the small- and large-scale structure. The nucleus
is faint enough that distortions of the nuclear PSF as a result of
pileup can be neglected. Although the values of $\beta$ are not well
constrained (values $>0.7$ are preferred) we find a good fit ($\chi^2
= 7.3$ for 7 d.o.f.) with a combination of a point source (for the
central active nucleus), a small-scale $\beta$ model with $\beta =
0.9$, $\theta_c = 0\farcs5$, representing the environment close to the
host galaxy, and a large-scale $\beta$-model with $\beta = 0.9$,
$\theta_c = 50'' \pm 10''$ ($r_{\rm c} = 75 \pm 7$ kpc), representing
the group. The radial profile and best-fitting models are shown in
Fig.\ \ref{285-profile}. The number of counts in the entire $\beta$
model (i.e. assuming spherical symmetry, neglecting the fact that some
material will have been removed from the lobes, and correcting for the
contribution of the $\beta$ model to the background region) is $860
\pm 90$ 0.5--5.0 keV counts: this number is insensitive in our fits to
the exact values of $\beta$ and $\theta_c$ used. Converting to a
luminosity using the temperature fitted above, this would correspond
to an undisturbed group bolometric X-ray luminosity of $(1.9 \pm 0.1)
\times 10^{42}$ erg s$^{-1}$ (note that this only includes statistical
errors, and takes no account of systematics in our assumptions) which
would place the system close to the temperature-luminosity
relationship for groups in general \citep[e.g.,][]{op04}, given the
relatively large uncertainty in temperature. We therefore feel
justified in assuming that this faint extended component is the group
environment in which 3C\,285's host currently resides. We comment
later in the paper (\S\ref{impact}) on the effect the current radio
source is likely to have on this environment.

The temperature determined above for the group-scale gas is clearly
cooler than that of the ridge (within the joint 90\% confidence limits)
for our choices of a fixed abundance of 0.35 solar, which might
suggest that the ridge and group-scale gas have a physically different
origin. However, if the two components are fitted jointly, each with
an independent normalization and temperature, lower abundances are
preferred ($<0.15$ at 90\% confidence), and for abundances fixed at a
low value the best-fitting temperature of the ridge emission comes down
to $0.9\pm0.15$ keV, consistent with that of the group gas within the
errors: accordingly, acceptable, though poorer, fits are possible to
the joint dataset with a single gas temperature around $0.9$ keV with
these low abundances. Although we adopt the temperatures determined
from our original fits in subsequent analysis, we cannot draw any
strong conclusions from the apparent difference in temperature between
the two components.

\subsubsection{3C\,442A}
\label{tail-fit}

\begin{figure}
\plottwo{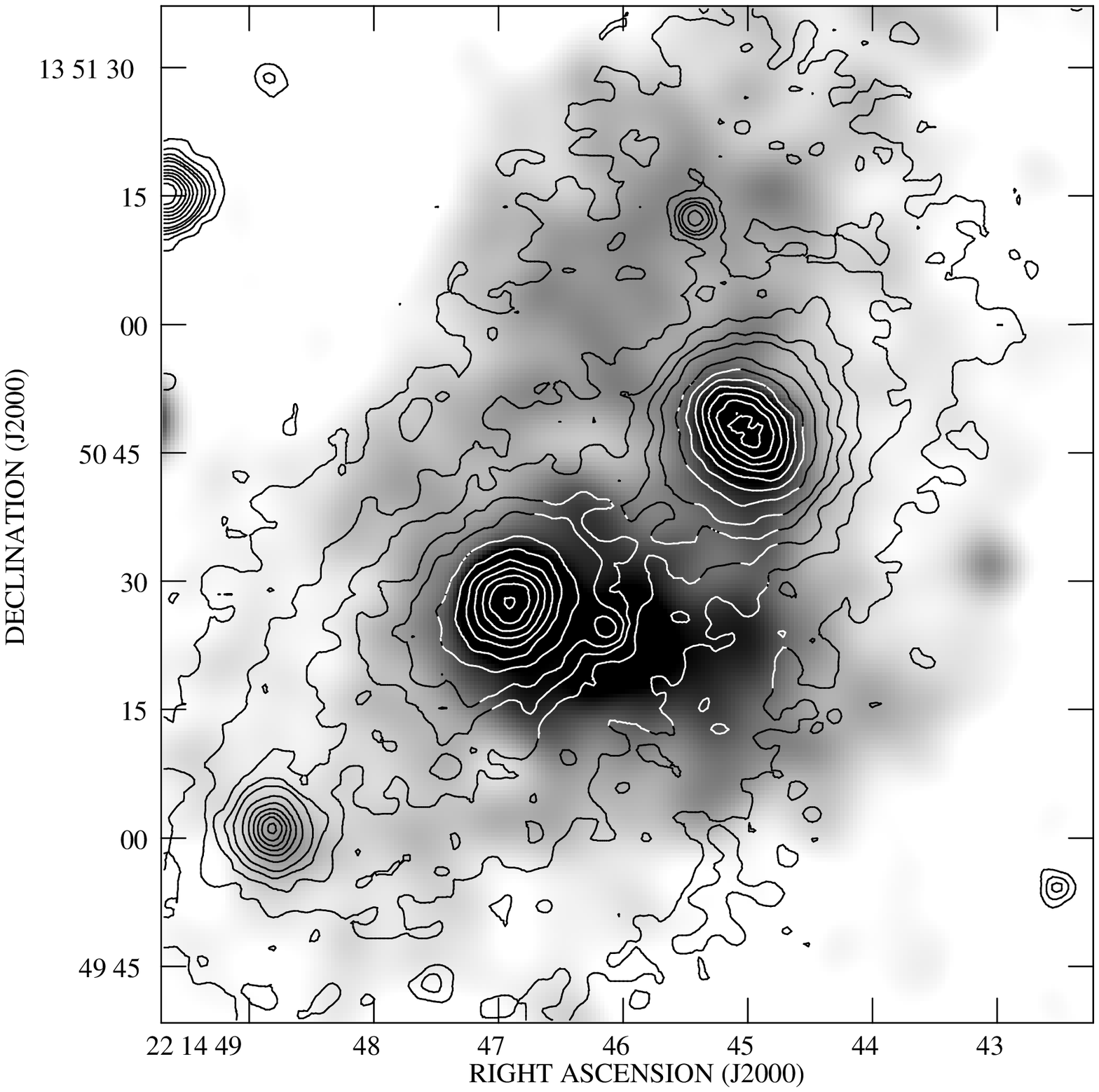}{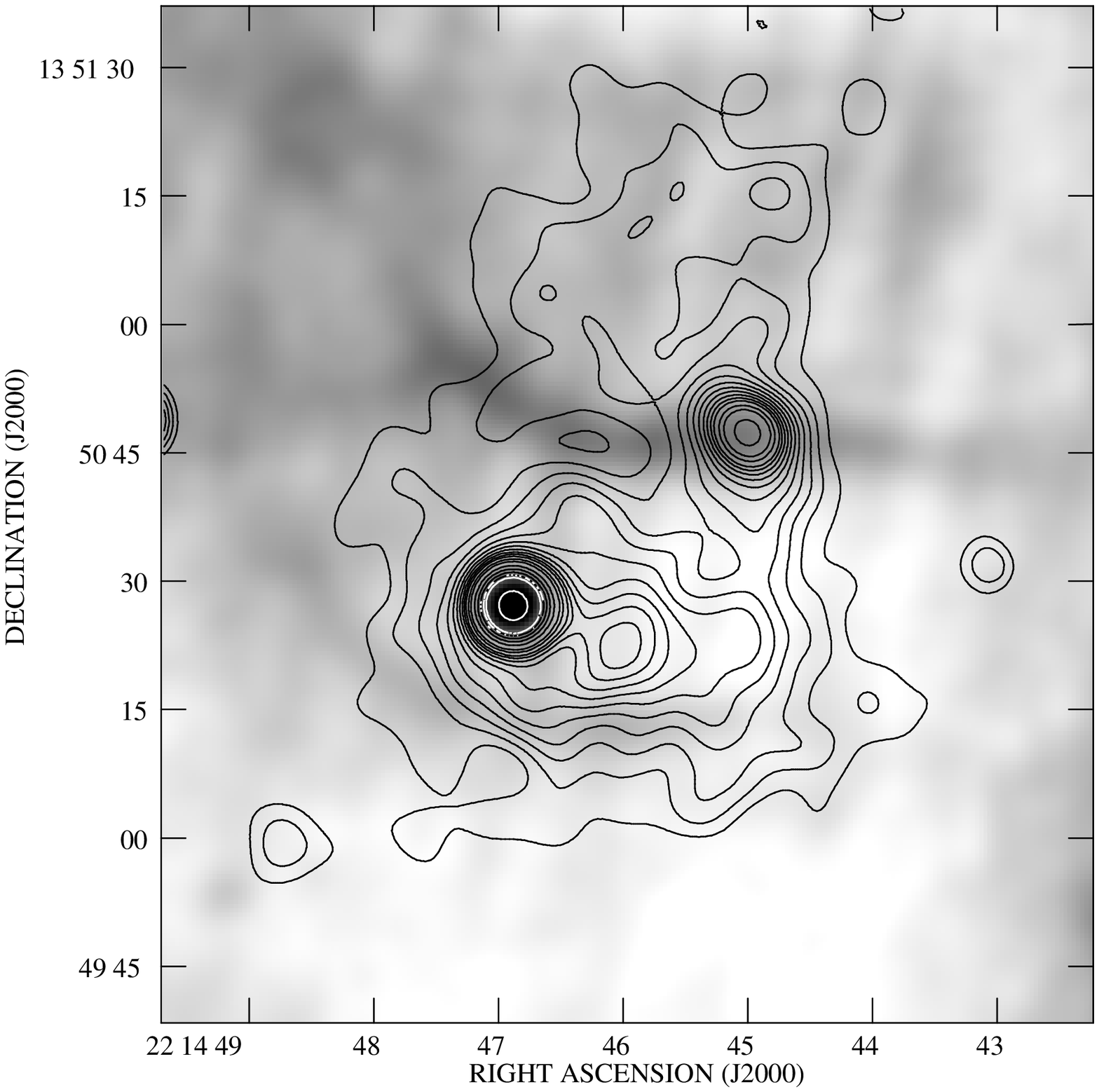}
\caption{Two views of the central X-ray emission from 3C\,442A. The
  X-rays shown here in both cases are the 0.5-5.0 keV events from all
  observations smoothed with a FHWM$=6''$ Gaussian. Left: X-ray in
  greyscale (black is 0.35 counts per {\it Chandra} pixel) with
  overlaid linear contours from the DSS2 red image, showing NGC 7236
  (top), the host galaxy NGC 7237 (middle) and their common optical
  envelope, including another small galaxy (bottom). Right: X-ray
  contours (at $0.12 \times (1, 1.3, 1.6, \dots, 3.7, 4.0, 4.5, 5.0,
  6.0,\dots 10.0, 15.0, \dots)$ counts pixel$^{-1}$: for clarity the
  lowest contour is well above the $3\sigma$ level) overlaid on a
  greyscale from the $5\farcs 37 \times 4\farcs 85$ resolution 1.4-GHz
  radio map (black is 3 mJy beam$^{-1}$). The images both show a
  spatial region of size $29 \times 34$ kpc.}
\label{442-centre}
\end{figure}

3C\,442A shows striking X-ray structure related to the host
interacting galaxy pair NGC 7236/7237 (Fig.\ \ref{442-centre}). Both
galaxies appear to have `tails' of extended, asymmetrical
X-ray-emitting material, presumably radiating thermally (see below),
with size scales around $40''$ (20 kpc). The tail of NGC 7237 lies in
position angle $\sim 250^\circ$ (N through E) and that of NGC 7236 in
position angle $\sim 45^\circ$. The tail of NGC 7237 was seen with the
{\it ROSAT} HRI by \citet{hw99}. As there is no evidence for
corresponding emission in the optical image (Fig.\ \ref{442-centre}),
where the distorted isophotes of the two galaxies are on smaller
scales \citep{bh88}, the tails must be formed not by tidal processes
but by some hydrodynamical process such as ram pressure stripping
\citep[e.g.,][]{asps03} or Bondi-Hoyle wake formation \citep{s00}. The
X-ray tails are in the direction expected from the compression of the
outer optical isophotes, which suggests that NGC 7237 is currently
moving towards the NE and NGC 7236 towards the SW \citep{bh88}.
Equally striking is the relationship between these tails and the
filamentary structure in the radio, as seen in the right-hand panel of
Fig.\ \ref{442-centre}. The brightest filament in the center of
3C\,442A passes directly through the gap between the two tails
(crossing the nucleus of NGC 7236), while the other bright filament in
the system runs along the S boundary of the NGC 7237 tail. Given this,
together with the qualitative similarity between the shape of the NGC
7237 tail (E-W at the W end, curving to NE-SW at the N end) and those
of the two filaments, it seems almost certain that the filamentary
structure arises from interaction between the radio plasma and the
tails. We return to this point below (\S\ref{tails}).

We extracted X-ray spectra for the two tails using rectangular regions
with adjacent background (excluding emission from the nuclei of both
galaxies), and fitted them with MEKAL models with free abundance. This
gives a good fit to the NGC 7237 tail ($\chi^2 = 51.8$ for 56 d.o.f.,
$kT = 0.77 \pm 0.04$ keV, abundance $0.3_{-0.1}^{+0.2}$ solar) and an
acceptable though rather poor fit to the NGC 7236 tail ($\chi^2 =
77.9$ for 44 d.o.f., $kT = 0.87_{-0.05}^{+0.15}$ keV, abundance $0.2\pm
0.1$). The bolometric X-ray luminosity of the tail regions with these
fits is $2.6 \times 10^{41}$ erg s$^{-1}$ and $2.1 \times 10^{41}$
erg s$^{-1}$ respectively.

\begin{figure}
\plotone{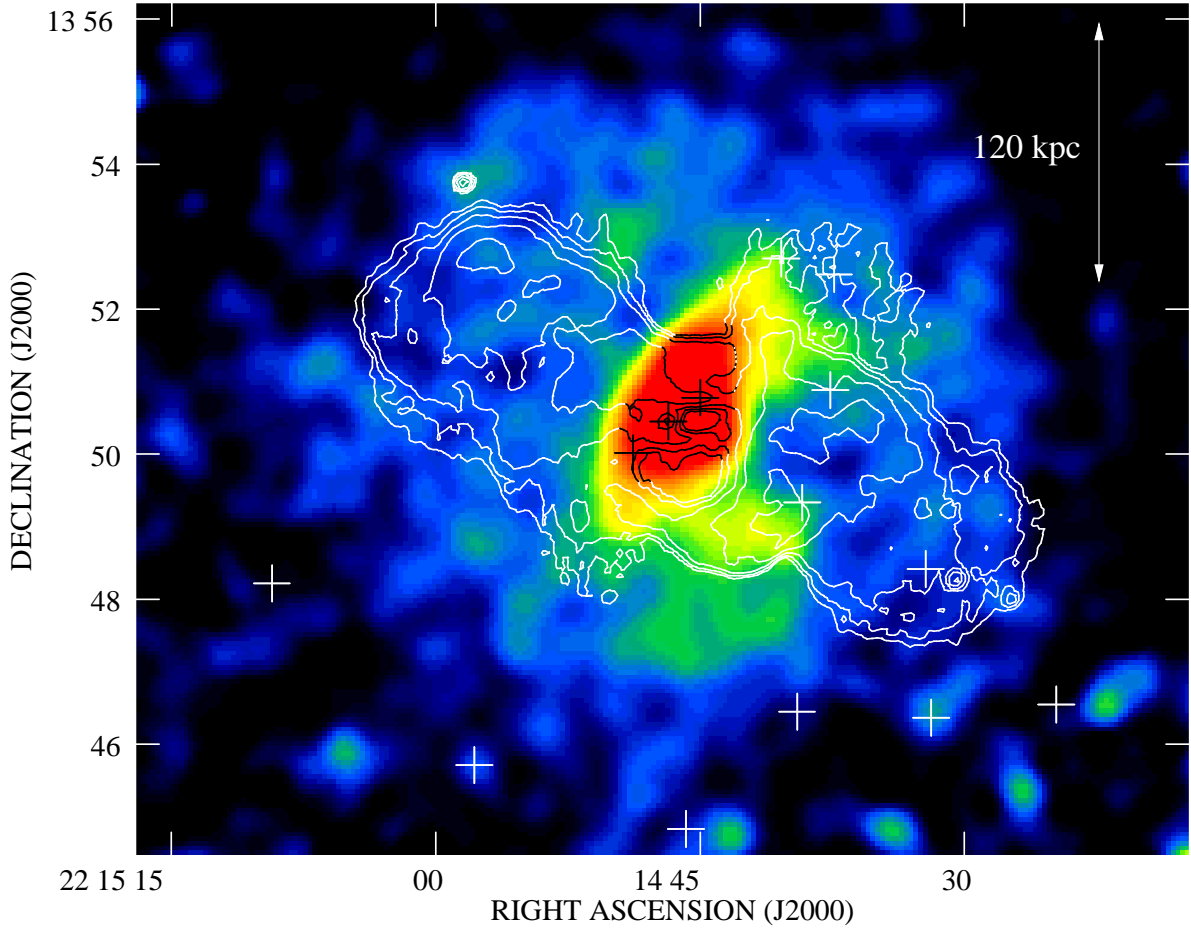}
\caption{Large-scale structure in the environment of 3C\,442A. Colors
  show the combined datasets in the energy range 0.5-5.0 keV after
  point source removal, weighted combination using exposure maps, and
  smoothing with a FWHM$=30''$ Gaussian. Excess noise in the bottom
  right of the image is present because this region was covered by
  fewer of the {\it Chandra} observations. Overlaid are contours of a
  1.4-GHz radio map of 3C\,442A at $7\farcs5$ resolution, at $0.25 \times
  (1,2,4,\dots)$ mJy beam$^{-1}$. Crosses mark the positions of nearby
  group galaxies, from Table \ref{442a-sdss}.}
\label{442-large}
\end{figure}

On larger scales, the tails merge into an elongated region that is
very similar in appearance to the `ridge' of 3C\,285, confirming the
result of \citet{hw99}. To illustrate this we masked out all point
sources (including the central nuclear point sources) and produced an
exposure-corrected image from the combined dataset, extending to the
edges of the combined field of view of the observations. Smoothing
this with a large Gaussian shows the large-scale structure in the
source (Fig.\ \ref{442-large}). The `ridge' in 3C\,442A extends
perpendicular to the radio source axis, as in 3C\,285, for 4.5 arcmin
(150 kpc), again (as in 3C\,285) terminating near a bright galaxy that
is probably a group member. Excess X-ray emission extends to the N and S of
the SW lobe and there is a clear deficit of emission coincident with
the lobe itself (and hints of a deficit associated with the NE lobe as
well). We extracted a spectrum for the ridge using a $2\farcm6 \times
1\farcm4$ rectangle centered on NGC 7237, excluding only the nuclear
point sources of the two galaxies. This spectrum clearly contains much
of the emission of the two X-ray tails discussed above. Although the
spectrum is roughly consistent with the type of model that might have
been expected from the results of fitting to the tails, the data are
not very well fitted with a single-temperature model ($\chi^2 = 194$
for 146 degrees of freedom: $kT = 0.95\pm 0.04$ keV, abundance $0.26
\pm 0.06$ solar) and the fit is improved if a two-temperature model is
used ($\chi^2 = 173$ for 145 d.o.f., $kT_1 = 0.71_{-0.18}^{+0.12}$
keV, $kT_2 = 1.14_{-0.14}^{+0.27}$ keV: here we fix abundance to 0.3
solar, as determined in the fits above, since it is degenerate with
relative normalization of the MEKAL models). This may suggest that the
`ridge' is a mixture of cool gas from the tails and hot gas from
another source. The bolometric X-ray luminosity of the ridge in these
models is $(8 \pm 2) \times 10^{41}$ erg s$^{-1}$.

Azimuthally symmetrical emission on scales of $10'$ is also clearly
visible in the image of Fig.\ \ref{442-large}. As with 3C\,285, this
seems likely to be thermal emission from a typical group-scale
atmosphere. We extracted spectra from a $4\farcm5$ radius source
circle around the source, with background taken from an adjacent
annulus, using the three of the four observations (OBSIDs 5635, 6353
and 6392) with enough of a distance between the source and the edge of
the chips to allow this. Point sources and the rectangular region
around the `ridge' were excluded, but the lobes were not, as there is
no direct evidence that their X-ray emission is significant. Counts
from the background dominate the spectrum (there were $\sim 2700$ net
counts but $\sim 18000$ total counts in the extraction region) and the
spectra were heavily binned before fitting to obtain adequate signal
to noise. Fitting with a MEKAL model we obtain a reasonable fit
($\chi^2 = 43.6$ for 39 d.o.f.) but with a poorly constrained
temperature, $kT = 2.0_{-0.5}^{+1.5}$ keV. Abundance is not
constrained in these fits and we set it to 0.3 solar, matching our
results for the ridge. The temperature we determine here is
appropriate for a rich group or poor cluster (it is consistent, for
example, with the Virgo cluster at $kT = 2.3$ keV).

\begin{figure}
\plotone{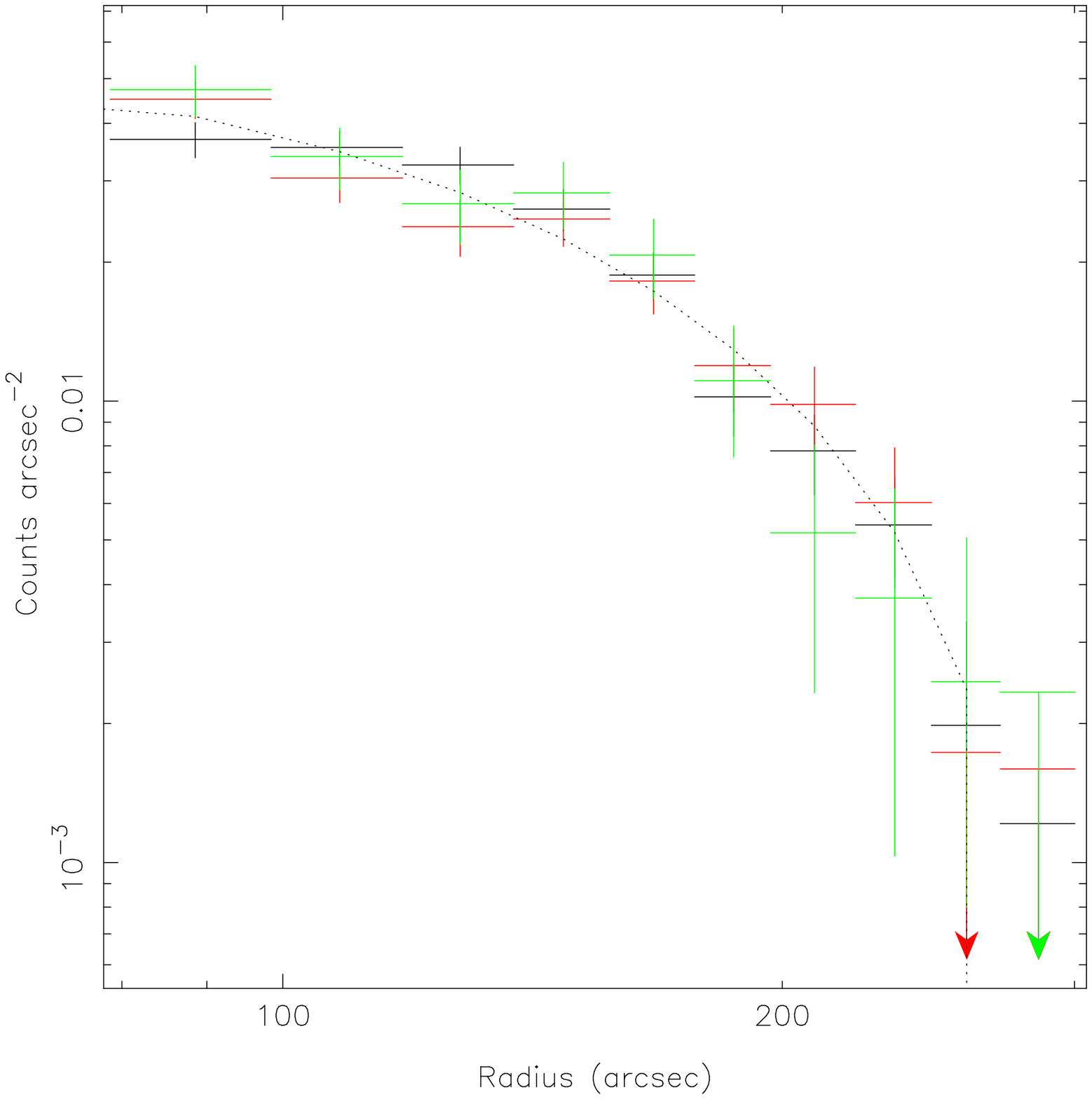}
\caption{Background-subtracted radial profile of extended emission in
  3C\,442A within $270''$ (135 kpc) after masking of `ridge' structure. The three
  profiles are taken from each of the three observations discussed in
  the text: the dotted line is the best-fitting $\beta$ model. The
  radial profiles have been renormalized to contain the same
  integrated number of counts.}
\label{442-profile}
\end{figure}

We extracted a radial profile for the large-scale emission, again
using the three OBSIDs for which annuli out to $5'$ could be fitted on
the ACIS-I chips and correcting for vignetting using exposure maps.
All structure on scales less than $80''$, as well as larger-scale
emission from the `ridge' and point sources, were excluded from the
profile. We fitted $\beta$ models to the profile, and although the
restricted angular range of the profile meant that it was not possible
to constrain $\beta$ ($\beta$ values $>0.7$ are preferred) we obtained
a good fit ($\chi^2 = 21.6$ for 27 d.o.f.) with $\beta = 0.9$,
$\theta_c = 220'' \pm 20''$ ($r_{\rm c} = 110 \pm 10$ kpc; error for
fixed $\beta$ value only). Fig.\ \ref{442-profile} shows the profile
and best-fitting $\beta$ model. Our fits allow us to estimate the
0.5-5.0 keV count rate in the three observations used, extrapolating
the $\beta$ model, as $0.18 \pm 0.02$ s$^{-1}$, which, assuming the
temperature fits above, implies a bolometric X-ray luminosity of $(4
\pm 0.5) \times 10^{42}$ erg s$^{-1}$ (again, errors quoted here come
from the uncertainty in the radial profile alone and do not include
systematic effects). Given the large uncertainty on derived
temperature, this is consistent with the temperature-luminosity
relation for the high-$kT$ end of the group population (or low-$kT$
end of the cluster population) particularly if the true temperature is
closer to the lower end of its 90 per cent confidence range. A lower
temperature would be more probable given the velocity dispersion for
the group calculated in \S\ref{442-vd}, but we note that groups around
radio galaxies of similar luminosity to 3C\,442A often have
temperatures that exceed those predicted from the
temperature-luminosity relation \citep{chb05a}.

\section{Discussion}

In this section we first consider (\S\ref{dynamics}) the implications
of these observations for the dynamics of the radio sources. We then
go on to consider the origin of the asymmetrical `ridge' structures in
the two systems, and show (\S\ref{origin}) that it is unlikely that
the radio galaxy can have given rise to these features, implying that
the environment shapes the radio source, rather than vice versa. This
allows us to construct simple pictures of the history of both systems
in terms of the interactions of gas and radio sources. The impact of
the radio sources on their group environments may be substantial, and
we comment on this in (\S\ref{impact}). The process of ridge formation
and interaction may be important to other sources with central gaps in
the radio lobes, and we discuss these in \S\ref{gaps}. Finally, in
\S\ref{tails}, we discuss the gas being stripped from the merging
galaxies hosting 3C\,442A and its interaction with the filamentary
radio structure at the center of the source.

\subsection{Dynamics of the radio sources}
\label{dynamics}

To discuss the effects of the X-ray emitting material on the radio
source it is first necessary to estimate the pressure in the radio
lobes. We assume initially that only the radiating electrons/positrons
and magnetic field contribute to the internal pressure: it follows
(since the sound speed in a relativistic plasma is high) that it is
appropriate to quote a single value for the internal pressure. Minimum
pressures can then be derived in the standard way for the lobes of
both sources given their sizes and flux densities. We use an electron
spectrum with $\gamma_{\rm min}=10$ and $\gamma_{\rm max} = 10^5$,
where $\gamma$ is the Lorentz factor of the electrons ($\gamma =
E/m_{\rm e}c^2$): the results are insensitive to the precise choices
of the low- and high-energy cutoffs. The electron spectrum has a
broken powerlaw form, such that $n_e \propto \gamma^{-p}$ for $\gamma
< \gamma_{\rm break}$ and $n_e \propto \gamma^{-(p+1)}$ for $\gamma >
\gamma_{\rm break}$. $p$, the low-energy electron energy power-law
index (`injection index') is taken to be 2.0 and the value of
$\gamma_{\rm break}$ is determined from a fit to the observed radio
lobe spectrum. The filling factor is set to unity (as standard for
minimum-energy calculations) and the magnetic field is assumed to be
tangled on small scales. For 3C\,285, the possible detection of
inverse-Compton emission via scattering of CMBR photons from the N
lobe gives us a {\it measurement} of the pressure, again assuming that
only radiating particles and field are present, since it gives a
direct constraint on the number density of electrons, which in turn
fixes the magnetic field strength given the observed synchrotron
emissivity. For 3C\,442A, the non-detection of inverse-Compton
emission gives us limits on the extent to which the lobe pressures can
be dominated by electrons.

We determine pressures as a function of radius in the groups for the
detected large-scale thermal components using the method of
\citet{bw93}, assuming spherical symmetry. For the
`ridge' components the appropriate assumptions are less obvious. We
derive approximate pressures on the assumption of uniform particle
density, converting from the {\sc xspec} MEKAL normalization to
density and thus pressure in the standard way. We use the two measured
dimensions of the spectral extraction region for the ridge to estimate
its size in the plane of the sky. The extension along the line of
sight is uncertain, and we consider two cases, treating the ridge either
as a cylinder whose long axis is in the plane of the sky, or as a disk
viewed edge-on. Because the density depends only on the square root of
the volume, these two geometries do not give rise to markedly
different densities and pressures.

\subsubsection{3C\,285}

The minimum pressures in the NE and SW lobes of 3C\,285 are $1.0
\times 10^{-13}$ Pa and $0.5 \times 10^{-13}$ Pa respectively (the
difference arises because of the fainter flux density and larger
volume of the SW lobe). If we assume that all the X-ray flux from the
NE lobe is inverse-Compton in origin, its pressure is $(3 \pm 1)
\times 10^{-13}$ Pa, where the error comes from the uncertainty on the
measured X-ray flux density. This is an upper limit on the pressure if
the lobes are electron-dominated.

For the group-scale emission, the extrapolated pressure at the base of
the lobes, $\sim 10''$ (15 kpc) in projection from the nucleus, using
the best-fitting $\beta$ model, is $(4.0 \pm 1.5) \times 10^{-13}$ Pa
(the dominant error here is the uncertainty on the best-fitting X-ray
temperature: the uncertainties on density from the $\beta$ model fits
are far smaller). At the far end of the lobes, $\sim 90''$ (140 kpc)
from the core, the thermal pressure is $(6 \pm 2) \times 10^{-14}$ Pa.
Formal errors due to the arbitrarily assumed abundance of 0.35 solar
cannot be estimated from the data, but pressure would increase by a
factor 1.5 if the abundance were 0.1 solar and decrease by a factor
1.5 if it were 1.0 solar, giving a range which is similar in magnitude
to that derived from the uncertainties in temperature. Even with the
large uncertainties on both internal and external pressure, it is
clear that 3C\,285 resembles other low-power FRII sources in poor
environments \citep[e.g.,][]{cbhw04} as well as more powerful objects
\citep[e.g.,][]{hbch02} in having lobes that are close to being in
pressure balance with the external thermal environment. It seems
likely that at their far ends the lobes are continuing to expand at
subsonic or mildly supersonic speeds (even the maximum possible
external thermal pressure only just balances the internal minimum
pressure of the NE lobe here), while at the ends closest to the
nucleus they can no longer expand transversely.

The pressure in the central `ridge', assuming our best-fitting
temperature and a uniform distribution of X-ray emitting gas, is $(4.6
\pm 0.9) \times 10^{-13}$ Pa (cylindrical geometry) or $(3.1 \pm 0.6)
\times 10^{-13}$ Pa (disk geometry), where the errors are again
dominated by the errors on our spectrally determined temperature (we
use the best-fitting single-temperature model for simplicity) and
again extreme variations in abundance could cause pressure variations
by a factor 1.5 in either direction. This ridge pressure is in some
sense a mean pressure in the region --- if there is a density
gradient, as the surface brightness suggests, then the pressure will
be higher closer to the center and lower at the edges. Thus, the fact
that the `ridge' pressure is comparable to or higher than the lobe
pressure means that if our radio pressure calculations are correct
(i.e. the lobes are not dominated in pressure terms by some
non-radiating component such as protons) then we would predict that
the lobes must avoid the region of the ridge. This is consistent with
the observation (Fig.\ \ref{285-smooth}) that the lobe plasma largely
avoids the ridge; in fact, the extensions of the lobes at the base to
the S (`wings') are on scales that suggest that the lobe plasma may be
flowing around the ridge. The anticoincidence between the lobe material
and the ridge is not perfect, but this is likely to be due to
projection, since the jet-counterjet asymmetry in 3C\,285 strongly
suggests that the radio source is not precisely in the plane of the
sky.

\subsubsection{3C\,442A}

For 3C\,442A we measure minimum pressures (using the assumptions
discussed above) from the outer regions of
the lobe only, ignoring the central region where an assumption of
uniform plasma filling factor would clearly not be appropriate, so
that the pressures measured here are for the parts of the source that
lie outside the ridge region on Fig.\ \ref{442-large}. The minimum
pressures we obtain are $2.1 \times 10^{-14}$ and $1.4 \times
10^{-14}$ Pa for the NE and SW lobes respectively. These are lower
than the 3C\,285 pressures, since 3C\,442A has both lower radio
luminosity and a slightly larger physical size.

The external thermal pressures are much larger than the minimum
pressures at all points in the radio source. At $50''$ (25 kpc), the
inner edge of the region measured for the NE lobe, the extrapolated
pressure (our $\beta$-model fit does not extend to this radius) is
$1.0_{-0.25}^{+0.75} \times 10^{-12}$ Pa (where the large (90\%
confidence) errors we quote come entirely from the poorly constrained
temperature: the uncertainties from $\beta$-model fitting are
negligible by comparison, though there is a possibility of systematic
errors due to the extrapolation, and we assume that large deviations
from the abundance fitted to the ridge are not likely). At the edges of
the lobes at $4\farcm5$ -- $5\farcm5$ (150--180 kpc) the pressures are
still $3_{-0.75}^{+2.25} \times 10^{-13}$ Pa, at least an order of
magnitude above the radio minimum pressures. Thus, although 3C\,442A's
radio structure is not that of a classical FRI radio source, it obeys
a tendency that has been known for many years (e.g.,
\citealt*{mfgh88,kbe88,fsgg90,tpik90,fpf92,bvfe93,wbc95,hwb98,wb00,chbw03};
\citealt{hsw05})
and rediscovered recently in the context of cluster-center sources
\citep[e.g.,][]{dft05} for low-power radio sources to have minimum
pressures that lie substantially (often 1--2 orders of magnitude)
below the external thermal pressure.

Radio lobes cannot evolve to a state in which they are substantially
underpressured with respect to the external medium: both theory
\citep[e.g.,][]{ka97} and observation \citep{kvfj03} imply that they
are likely to be strongly overpressured initially, and, while the
internal pressure should drop as the linear size of the source
increases, any drop below the external pressure will cause the
external thermal material to expand into the lobe region, compressing
the lobe plasma until pressure balance is restored. Observations that
suggest that the minimum pressure is below the external pressure
therefore imply some additional contribution to the internal pressure,
via a large departure from equipartition, a low filling factor, and/or
a dominant contribution from non-radiating particles such as protons.
A departure from equipartition in the direction of electron dominance
is almost certainly ruled out, since at pressure balance it would
produce a 1-keV flux density via inverse-Compton scattering of the
CMBR of about 90 nJy per lobe, giving a count rate in {\it Chandra}
observations that would exceed that from the thermal emission from the
entire group. In fact, if anything there is a deficit of X-ray
emission from the lobes, particularly the W lobe, as mentioned above.
As magnetic domination by large factors is not observed in lobes of
sources where inverse-Compton emission can be studied \citep{chhb05b},
we favor a model in which the lobe pressure is dominated by
non-radiating particles. In the case of more typical FRI sources it
has been suggested that protons from the external thermal material
could be entrained and efficiently heated to high energies by the jets
(temperatures much greater than that of the external thermal material
are required to avoid the detection of this component through its
thermal bremsstrahlung). As no jets of the sort seen in normal FRIs
are present in 3C\,442A, this explanation would only be viable if the
lobes were at one time driven by FRI-type jets that are now no longer
present, or no longer visible. Alternatively, a population of
relativistic protons could have been injected into the lobes along
with the radiating electrons (though this is inconsistent with the
picture in FRII lobes where an energetically dominant proton
population is disfavored: \citealt{chhb05b}) or the lobes could have
engulfed large regions of thermal material, though again some unknown
efficient heating mechanism would be required.

The pressure in the `ridge' in 3C\,442A, again assuming uniform density,
is $(1.1 \pm 0.1) \times 10^{-12}$ Pa (cylindrical geometry) or $(8.1
\pm 0.1) \times 10^{-13}$ Pa (disk geometry), so, unsurprisingly, the
ridge pressure is equal to or exceeds the extrapolated central pressure
of the extended emission, in spite of the lower temperature of the ridge
material (here again we are using the best-fitting single-temperature
model for the ridge). The pressures derived by \citet{wkbh07},
using a $\beta$-model fit to the ridge region, are similar. Thus even if
the lobe is in pressure balance with the thermal material further out
(as it must be, to avoid collapse on a sound-crossing time) the lobe
pressure will be lower than the ridge pressure, and the lobe material
will be excluded from the ridge region unless it can achieve higher
pressures. Fig.\ \ref{442-large} shows that material in the lobes,
particularly the W lobe, does seem to be avoiding the ridge and to be
being pushed out in wings to the N and S, as well as possibly along
the line of sight (note the strong radio brightness gradient to the W
of the ridge). There is still radio-emitting material in the ridge region,
but, as discussed above, the brightest components are in the form of
filaments that are likely to be interacting strongly with the
small-scale material from the galaxies, and these filaments will have
pressures substantially higher than the lobe material as a whole (see
below, \S\ref{tails}, for more discussion of this point).

\label{ridge-pressure}

\subsection{Origin of the `ridges' and the evolution of the radio galaxies}
\label{origin}

Is the `ridge' structure intrinsic to the system, or could it be a
result of the interaction of the radio galaxy with an initially
spherically symmetrical gas distribution? There are two strong
indications that the former is the case: firstly, the ridges align
well with the local distribution of galaxies and starlight, which
cannot be affected by the radio galaxy, in both 3C\,285 and 3C\,442A;
secondly, in 3C\,442A, we observe material leaving the two central
bright galaxies via the tails and merging with the ridge.
Quantitatively, we can compare the mass in the ridges (or rather in
the somewhat smaller spectral extraction regions used to derive
temperatures and pressures) with the masses of thermal material
affected by the lobes. Using the approximations of uniform density
from the previous section, we find that the `ridge' gas masses are in
the range $2$ to $3 \times 10^{10} M_\odot$ for 3C\,285 and $1$ to
$1.5 \times 10^{10} M_\odot$ for 3C\,442A. Using the $\beta$-model
fits to the larger-scale thermal emission, we find that for 3C\,285
the ridge mass is similar to, but somewhat greater than, the total
mass of gas that would be present in the lobe regions if the lobes
were not there: thus, since only a fraction of the displaced gas would
be expected to end up in the ridges, it is essentially impossible for
the radio lobes to be responsible for producing the ridge in 3C\,285
(though of course here we have to assume that the distribution of the
gas in the regions away from the lobes of the radio source now is
similar to its undisturbed distribution). The constraints are less
strong for 3C\,442A, with larger lobes, a denser thermal environment,
and a smaller ridge mass: roughly $1/3$ of the displaced gas would be
sufficient to produce the observed ridge. However, we know that in
3C\,442A at least some of the gas in the ridge did {\it not} come via
this route, since some of it must have passed through the tails. {\it
Some} contribution from compressed, displaced gas would help to
explain the overdensities seen to the N and S of the W lobe
(\S\ref{tail-fit} and Fig.\ \ref{442-large}) (although note that these
are not included in our mass calculations). We conclude that the ridge
structures in these two sources are largely or wholly independent of
the presence of the radio lobes.

Our outline models of the evolution of the two radio sources are
somewhat different, because we know that the formation of the ridge in
3C\,442A is still going on given the observations of tails of gas from
the two central galaxies (see \S\ref{tails}, below) while there is no
evidence that there is ongoing gas stripping in 3C\,285. 3C\,285 thus
presents the simpler situation, and we consider it first. In this
system we consider it entirely possible that the ridge was present in
something like its current configuration before the radio activity
began. It is possible, although the data do not require it, that some
of the gas that now forms the ridge was stripped from the galaxies that
merged to form the current host galaxy of 3C\,285. Whether this is
true or not, we have seen that the ridge is in rough pressure
equilibrium with the external group-scale gas, at least in the center
of the group. The timescale for the relaxation of the ridge back towards
a spherically symmetrical situation, assuming that the gravitational
potential is spherically symmetric, is thus likely to be comparable to
the sound-crossing time for the {\it long} axis of the ridge, i.e. a few
$\times 10^8$ years. If the gravitational potential is non-static, the
relevant timescale is the gravitational relaxation time, whch could be
longer still. Since the expansion of the 3C\,285 lobes is likely to
have been supersonic initially (and may still just be supersonic at
present) the growth of the lobes to similar size scales can take place
in a much shorter time, a few $\times 10^7$ years, and so it seems
consistent to suppose that the ridge has not evolved greatly while the
lobes have expanded. The expanding lobes would initially be confined
by ram pressure and so would expand preferentially into the
lower-density group-scale medium. Later, as the lobe internal pressure
dropped, the ridge material would push any lobe material out of the
center of the group, leading to the situation we now see, where the
currently observed ridge is in rough pressure balance with lobes that
avoid the central part of the system.

The more complex situation in 3C\,442A is discussed in detail by
\citet{wkbh07}. In outline, our suggested picture is that here
a large, old radio source could well have been present {\it before} the
start of the ongoing interaction that produced the presently visible
ridge material. By stripping and disrupting the hot gas in the center of
the group, the start of the interaction/merger between NGC 7236/7 may
have terminated the energy supply to the lobes. In any case, the
resulting ridge has now largely driven the lobe material out of the
center of the group, although the interaction with radio filaments
(see \S\ref{tails} below) shows that this process is not yet complete.
Buoyancy will remove the remaining filaments on a relatively short
timescale (unless they are magnetically anchored to one or other of
the galaxies), while the ridge may continue to expand, driving the radio
lobes ahead of it, as the merger of the two galaxies proceeds to
completion.

\subsection{Effect of the radio galaxies on the groups}
\label{impact}

Crudely, the work done on the thermal material by the expanding radio
galaxy ($\int p{\rm dV}$) may be estimated as $\sim p_{\rm ext}V_{\rm
lobe}$, or $\sim p_{\rm int}V_{\rm lobe}$ for pressure balance. This
estimate probably gives constraints that are of the right order of
magnitude. For 3C\,285 the numbers are between $1$ and a few $\times
10^{52}$ J, depending on whether the source is in pressure balance or
mildly overpressured at the present time, while for 3C\,442A the
energy input assuming rough pressure balance at the radio lobes is
$\sim 5 \times 10^{52}$ J. In both cases the energy stored in the
lobes is 3 times this value (we assume a relativistic equation of
state). These numbers may be compared to the total energy (${3\over2}
NkT$) of the thermal material in the groups (neglecting the ridges)
integrated out to the radius of the edge of the lobes (150 kpc for
3C\,285, 160 kpc for 3C\,442A). For 3C\,285 this thermal energy is $(6
\pm 2) \times 10^{52}$ J: thus the energy input by the lobes is
certainly significant, and the total energy input by the source over
its lifetime will be sufficient to unbind or move to large radii a
large fraction of the currently existing group-scale gas. For
3C\,442A, with a denser medium and larger temperature, the thermal
energy of the gas is $3.2_{-0.8}^{+2.4} \times 10^{53}$ J, so that the
impact of the radio source is less relative to the existing thermal
energy than in 3C\,285: however, again, the total energy input, once
the internal energy of the lobes is thermalized, should be very
important. Spectral aging (via inverse-Compton losses) puts a limit on
the lifetime of 3C\,442A of $\sim 10^8$ years, while in 3C\,285 we
expect a lifetime of a few $\times 10^7$ years from the dynamical
arguments of the previous section. Thus in both cases the rate of
energy input is more than enough (by 1-2 orders of magnitude) to
offset the radiative cooling of the group-scale gas given the
bolometric luminosities determined in \S\ref{group-properties},
although in the case of 3C\,442A some of that work probably comes not
from the active nucleus but (via the expansion of the ridge) from the
dissipation of the energy of the galaxy-galaxy merger.

\subsection{Central gaps in radio sources and the presence of ridges}
\label{gaps}

A significant minority of radio galaxies, particularly low-luminosity
FRIIs, have radio lobes that are apparently sharply truncated at an
inner edge, as first pointed out by \citet{js76}. The truncation means
that in radio images there are well-defined sharp-edged gaps between
pairs of radio lobes that typically contain no lobe-related radio
emission. \citet{gw00} have compiled a sample of such objects, and
suggest that the central gaps between the lobes are a result of
hydrodynamical interactions with a {\it cold} thermal medium. While
this may be true in some cases, our results show that asymmetrical
structures in the hot phase of the IGM can have the same effect. A few
other radio galaxies with truncated lobes have X-ray observations, and
while some (e.g., 3C\,33, \citealt{kbhe07}; 3C\,227, Hardcastle \etal\
in prep.) show no obvious signs of asymmetrical X-ray structure on
scales comparable to the lobes, at least one other FRII radio galaxy,
3C\,401, has both elongated X-ray emission filling between the gap
between the lobes \citep*{rbs05}, and, crucially, some evidence for
elongation of the stellar population in the same E-W direction in {\it
Hubble Space Telescope} images \citep{csmp05}, strongly suggesting
that the `ridge' structure cannot simply be a result of compression of
pre-existing symmetrical gas in the cluster environment of the source.
Given the evidence that mergers are important in the environments of
FRII sources \citep[e.g.,][]{hsbv86} it is possible that central gaps
are related to interaction of the lobes with pre-existing or
developing X-ray structures in many other cases.

\subsection{The X-ray tails and the filamentary structure in 3C\,442A}
\label{tails}

`Tail' or `fan'-like X-ray structures have been seen in other strongly
interacting galaxy pairs. The twin tails in NGC 7236/7 (Fig.\
\ref{442-centre}) are very similar in structure and physical size to
those seen by \citet{mkjf06} in NGC 4782/3, which consists of two
interacting galaxies of similar mass (one hosting the radio source
3C\,278) with a projected separation of around 13 kpc. The tails in
the 3C\,278 system are overpressured with respect to the surrounding diffuse
medium, implying that they originate in ram pressure stripping of the
halos of the interacting ellipticals, and \citeauthor{mkjf06} carried
out a detailed analysis of the X-ray surface brightness around the
leading edge of one of the galaxy profiles to estimate the approximate
speed of motion of the galaxy relative to the IGM. Because both tails
in 3C\,442A have some excess X-ray emission in front of the galaxies,
we cannot repeat the \citeauthor{mkjf06} analysis for 3C\,442A, but we
can make a less rigorous estimate by determining the pressure inside
the tails. The tail of NGC 7237 has only a weak surface brightness
gradient over the inner 40$''$ (20 kpc) of its length, so we model the
X-ray emission as a uniform cylinder of length 20 kpc and radius 7
kpc. Our spectral fit to the tail (\S\ref{tail-fit}) then gives a
density of $1.1 \times 10^4$ m$^{-3}$ and a pressure of $(3.1 \pm 0.1)
\times 10^{-12}$ Pa (where the dominant uncertainty comes from the
small uncertainty on temperature). To explain why the two dominant
galaxies of the group are being ram-pressure stripped {\it now},
rather than having being completely stripped at a much earlier stage
of the group's evolution, we must assume that it is their current
close interaction that is responsible for the stripping: in this case
it is appropriate to compare their pressure and density to the
external medium in the ridge, which (as discussed above) we infer to be
made up primarily of previously stripped material, rather than the group-center
hot gas: in any case, the tails are clearly embedded in the ridge. This
comparison gives a density contrast of a factor 3.3--4.5 (depending on
the geometry assumed for the ridge, see \S\ref{ridge-pressure}) and a
pressure contrast of a factor $2.8 \pm 0.1$, assuming cylindrical
geometry since that gives the closest pressure balance with the
larger-scale medium. The tails in 3C\,442A are thus overpressured and
the leading edge must be ram-pressure confined, consistent with the
picture of \citeauthor{mkjf06} Following \citeauthor{mkjf06} in
adopting the relation of \citet*{vmm01} between pressure contrast and
Mach number $\cal M$, we find that this pressure contrast implies
${\cal M} = 1.2^{+0.1}_{-0.1}$, or a speed with respect to the ridge
material of $600\pm 50$ km s$^{-1}$. The models of merger dynamics by \citet{b88} imply
velocities $\sim 600$ km s$^{-1}$, in good agreement with this result.
Away from the leading edge, we expect the tails to expand transversely
into the medium of the ridge at the sound speed, giving rise to the short,
stubby structures that are observed and possibly to the apparent
flaring of the NGC 7236 tail (Fig.\
\ref{442-centre})

If the filaments seen in the radio image are directly interacting with
the tails, we expect that the pressure in the radio-emitting plasma
will be comparable to the pressures in the tail. The radio filaments
are marginally resolved in our highest-resolution radio images, with
radii of 1--2 arcsec. We can use observed radio surface brightness per
unit length to estimate that they would need to have a radius of
$0\farcs16$ (0.09 kpc) for the {\it minimum} pressures to be
comparable to the thermal pressure in the tails: this is probably
inconsistent with the radio data for the southern filament, and almost
certainly inconsistent with the data for the northern filament.
However, if the pressure in the filaments exceeds the minimum pressure
by the factor $\sim 10$ required in the lobes, then the radius must be
$\sim 1\farcs4$, 0.8 kpc, for pressure balance with the brighter tail,
which is similar to what we observe in the high-resolution radio
images. Thus it is plausible that the filaments are produced by direct
interaction with the tails, as suggested by their positions (Fig.\
\ref{442-centre}).

Assuming that there are not strong variations in the ratio of magnetic
field to electron energy density, the filaments must be overpressured
with respect to the lobes in order to be visible at all, and so must
be dynamically evolving structures. A model in which they are
associated with the galaxies, which we know to be moving with respect
to the radio plasma, is therefore {\it a priori} plausible. Since the
filaments lie to the sides of the X-ray tails (Fig.\ \ref{442-centre})
they cannot simply be a result of compression of the remaining central
radio-emitting plasma by the moving galaxies. It is plausible that
they arise in the shear layers between the stripped material in the
tails and the external gas.

\section{Summary and conclusions}

We have presented {\it Chandra} observations of the nearby radio
galaxies 3C\,285 and 3C\,442A. The main results of the paper can be
summarized as follows.

\begin{itemize}
\item The large-scale environments of both sources are groups,
  consistent with earlier optical results. The groups have similar
  X-ray luminosities, physical sizes and $\beta$-model parameters,
  although the best-fitting temperature for 3C\,442A's group is hotter. Both
  sources have elongated, 150-kpc scale excess X-ray emission which
  aligns with galaxies in the group, and which we refer to as the ridge.
  This excess X-ray emission is coincident with the regions in which
  there is little or no radio emission from the lobes.
\item Using temperatures and pressures derived from the X-ray
  observations, we have shown that the pressure in the ridges in both
  sources, within the errors, is equal to or exceeds any plausible
  internal pressure in the lobes. Thus the ridges are capable of
  driving, or having driven, the existing radio lobes out of the
  central regions of the source.
\item We have shown that there is direct quantitative evidence in the
  case of 3C\,285 (\S\ref{origin}), and strong qualitative evidence in
  the case of 3C\,442A, that the ridges are not the result of an
  interaction between pre-existing gas and the radio source, but must
  have arisen independently. In 3C\,442A, we see hot gas in the
  process of being stripped from the central interacting galaxies and
  merging into the larger-scale ridge, suggesting that the ridge has
  been formed by the stripping of the haloes of the two merging
  galaxies. In 3C\,285, the optical observations strongly suggest that
  the host galaxy has recently merged with a gas-rich system, and
  there is a tidal interaction with a more distant galaxy, although
  there is no evidence that the ridge is currently being fed with hot
  gas; it is thus possible (though by no means required by the data)
  that the ridge in 3C\,285 also has an origin in galaxy-galaxy
  interactions. We argue that interaction with similar elongated
  structures may be responsible for some of the observed central gaps
  in the lobes of other radio sources. A prediction of this picture is
  that there will be ridge structures in other groups that do {\it
  not} host large, powerful radio galaxies: so far few of these are
  known, but the group hosting NGC 5171 \citep{opf04} may be a good
  example.
\item Our detailed view of the tails of hot gas associated with
  3C\,442A allows us to estimate the Mach number and speed of NGC 7237
  (the host of 3C\,442A) with respect to the diffuse IGM. We obtain
  results consistent with detailed modelling of the merger dynamics
  \citep{b88}. The properties of the radio filaments at the center of
  3C\,442A are consistent with their being formed by interaction with
  the tails of the two galaxies in the system.
\item The X-ray nuclei of the two sources have spectra that are
  consistent with what is expected for objects of their emission-line
  type. No convincing evidence for X-ray counterparts to compact
  radio features, such as jets or hotspots, is found, and we have shown that
  the previously claimed optical counterpart to the E hotspot of
  3C\,285 is unlikely to be associated with the hotspot.
  X-ray emission is detected from the region of jet-induced star
  formation in 3C\,285 reported by \citet{vd93}.
\end{itemize}

\acknowledgements
We gratefully acknowledge financial support for this
work from the Royal Society (research fellowship for MJH) and NASA
(grants GO5-6100X and NAS8-03060). The National Radio Astronomy
Observatory is a facility of the National Science Foundation operated
under cooperative agreement by Associated Universities, Inc. This
research has made use of the NASA/IPAC Extragalactic Database (NED)
which is operated by the Jet Propulsion Laboratory, California
Institute of Technology, under contract with the National Aeronautics
and Space Administration. We thank an anonymous referee for helpful
comments which enabled us to improve the presentation of the paper.

{\it Facilities:} \facility{CXO (ACIS)}, \facility{VLA}

\end{document}